\documentclass[article]{aa} 
\usepackage{natbib}
\usepackage{graphicx}
\usepackage{txfonts}
\usepackage{longtable}

\begin{document}

\title{Evolution of the dusty infrared luminosity function from $z=0$ to $z=2.3$ using observations from \textit{Spitzer}}
\author{B. Magnelli\inst{1,2}
\and
D. Elbaz\inst{1}
\and
R.~R Chary\inst{3}
\and
M. Dickinson\inst{4}
\and
D. Le Borgne\inst{5,6}
\and
D. T. Frayer\inst{7} 
\and
C. N. A. Willmer\inst{8}}
\institute{Laboratoire AIM, CEA/DSM-CNRS-Universit\'e Paris Diderot, IRFU/Service d'Astrophysique, B\^at. 709, CEA-Saclay, F-91191 Gif-sur-Yvette C\'edex, France
\and
Max Planck Institut f\"ur extraterrestrische Physik, Postfach 1312, 85741 Garching, Germany\\
\email{Magnelli@mpe.mpg.de}
\and
Spitzer Science Center, California Institute of Technology, Pasadena, CA 91125, USA
\and
National Optical Astronomy Observatory, Tucson, AZ 85719
\and
UPMC Univ Paris 06, UMR7095, Institut d'Astrophysique de Paris, F-75014, Paris, France
\and
CNRS, UMR7095, Institut d'Astrophysique de Paris, F-75014, Paris, France
\and
National Radio Astronomy Observatory, P.O. Box 2, Green Bank, WV 24944, USA 
\and
Steward Observatory, University of Arizona, 933 North Cherry Avenue, Tucson, AZ 85721, USA}
\date{Received ??; accepted ??}


\abstract
{}
{
We derive the evolution of the infrared luminosity function (LF) over the last 4/5ths of cosmic time, using deep 24 and 70~$\mu$m imaging of the GOODS North and South fields.
}
{
We use an extraction technique based on prior source positions at shorter wavelengths to build the 24 and 70 $\mu$m source catalogs.
The majority ($93\%$) of the sources have a spectroscopic ($39\%$) or a photometric redshift ($54\%$) and, in our redshift range of interest (i.e., $1.3<z<2.3$) $\thicksim20\%$ of the sources have a spectroscopic redshifts.
To extend our study to lower 70 $\mu$m luminosities we perform a stacking analysis and we characterize the observed $L_{24/(1+z)}$ vs $L_{70/(1+z)}$ correlation.
Using spectral energy distribution (SED) templates which best fit this correlation, we derive the infrared luminosity of individual sources from their 24 and 70 $\mu$m luminosities.
We then compute the infrared LF at $z\thicksim1.55\pm0.25$ and $z\thicksim2.05\pm0.25$.
}
{
We observe the break in the infrared LF up to $z\thicksim2.3$.
The redshift evolution of the infrared LF from $z=1.3$ to $z=2.3$ is consistent with a luminosity evolution proportional to $(1+z)^{1.0\pm0.9}$ combined with a density evolution proportional to $(1+z)^{-1.1\pm1.5}$.
At $z\thicksim2$, luminous infrared galaxies (LIRGs: $10^{11} L_{\odot}<\,$L$_{\rm{IR}}$\,$ <10^{12} L_{\odot}$) are still the main contributors to the total comoving infrared luminosity density of the Universe.
At $z\thicksim2$, LIRGs and ultra-luminous infrared galaxies (ULIRGs: $10^{12} L_{\odot}<\,$L$_{\rm{IR}}$) account for $\thicksim49\%$ and $\thicksim17\%$ respectively of the total comoving infrared luminosity density of the Universe.
Combined with previous results using the same strategy for galaxies at $z<1.3$ and assuming a constant conversion between the infrared luminosity and star-formation rate (SFR) of a galaxy, we study the evolution of the SFR density of the Universe from $z=0$ to $z=2.3$.
We find that the SFR density of the Universe strongly increased with redshift from $z  = 0$ to $z = 1.3$, but is nearly constant at higher redshift out to $z = 2.3$.
As part of the online material accompanying this article, we present source catalogs at 24 $\mu$m and 70 $\mu$m for both the GOODS-North and -South fields.
}
{}
\authorrunning{Magnelli et al. }
\titlerunning{Infrared Luminosity Density at $0<z<2.3$}
\maketitle

\section{Introduction}
\indent{
The important contribution of infrared luminous galaxies (Luminous Infrared Galaxies, LIRGs: $10^{11} L_{\odot}<\,$L$_{\rm{IR}}$\,$ <10^{12} L_{\odot}$; Ultra-Luminous Infrared Galaxies, ULIRGs $10^{12} L_{\odot}<\,$L$_{\rm{IR}}$) in the evolution of the star-formation rate (SFR) history of the Universe is now well established up to $z\thicksim1$ \citep{chary_2001, franceschini_2001, xu_2001, elbaz_2002, metcalfe_2003, lagache_2004,lefloch_2005,magnelli_2009}.
Their contribution to the SFR density of the Universe increases with redshift up to $z\thicksim1$ where the bulk of the SFR density occurs in LIRGs.
Study of this evolution was made possible through the use of large and accurate spectroscopic and/or photometric redshift catalogs as well as deep 24 and 70 $\mu$m surveys obtained by \textit{Spitzer}.
\\}
\indent{
At $z>1.3$, the SFR history of the Universe has been derived by several studies \citep{perez_2005,caputi_2007} using deep 24 $\mu$m imaging and infrared bolometric correction estimated from local spectral energy distribution (SED) libraries \citep[][]{chary_2001,lagache_2003,dale_2002}.
All of these studies concluded that the relative contribution of ULIRGs to the SFR density of the Universe increases with redshift, and may even be the dominant component at $z\thicksim2$.
However, these conclusions still need to be confirmed since there are large uncertainties at high-redshift in transforming observed 24~$\mu$m flux densities to far-infrared luminosities \citep{papovich_2007,daddi_2007}.
To study the high-redshift evolution of the SFR density, one has to combine deep mid- and far-infrared observations in order to infer robust bolometric corrections and to clearly constrain the location of the break of the infrared luminosity function (LF).
\\ \\}
\indent{
At $z\thicksim2$, the observed 70 $\mu$m emission corresponds approximately to the rest-frame 24 $\mu$m luminosity, which was proven to be a good SFR estimator in the local Universe \citep{calzetti_2007}.
The reliability of this SFR estimator seems to hold at high-redshift since the SFR of $z\thicksim2$ galaxies estimated from their observed 70~$\mu$m flux densities and their radio emissions are in good agreement \citep{daddi_2007a}.
Thus, to get robust estimates of the SFR of distant galaxies, we decided to use deep 70~$\mu$m observations obtained by \textit{Spitzer}.
\\}
\indent{
The main difficulty of using 70 $\mu$m observations to study star-formation at $z\thicksim2$ resides in the limited depth of the existing \textit{Spitzer} data, even from the very deepest observations such as those in the Great Observatories Origins Deep Survey (GOODS).
In this study, we overcome this limitation by using a stacking analysis.
As shown in \citet{papovich_2007} this analysis allows characterization of the 24 vs 70~$\mu$m correlation and thus constrains the bolometric corrections to be applied to the 24~$\mu$m flux densities.
Using deep 24 and 70 $\mu$m images of the GOODS-North and South fields we find that the 24 vs 70~$\mu$m correlation observed at high redshift is significantly different from predictions by standard SED libraries.
This deviation can be interpreted as a possible signature of an obscured active galactic nuclei (AGN) \citep{daddi_2007} or simply as a SED evolution characterized by stronger polycyclic aromatic hydrocarbon (PAH) emission \citep{papovich_2007}.
Both interpretations are discussed and two different bolometric corrections are inferred.
Based on these bolometric corrections we derive the infrared LF in two redshift bins (i.e., $1.3<z<1.8$ and $1.8<z<2.3$).
For the first time these infrared LFs take into account the evolution of SED observed in high-redshift galaxies.
By comparison, extrapolation from the observed 24~$\mu$m emission alone results in significant overestimates in their infrared luminosity \citep{papovich_2007,daddi_2007} while extrapolations from the observed 850~$\mu$m emission constrains only the most luminous ULIRGs.
\\}
\indent{
Throughout this paper we will use a cosmology with $H_{0}=70\ \rm{km\ s^{-1}\ Mpc^{-1}},\Omega_{\Lambda}=0.7$ and $\Omega_{M}=0.3$.
}
\begin{figure*}
\centering
	\includegraphics[width=8.cm]{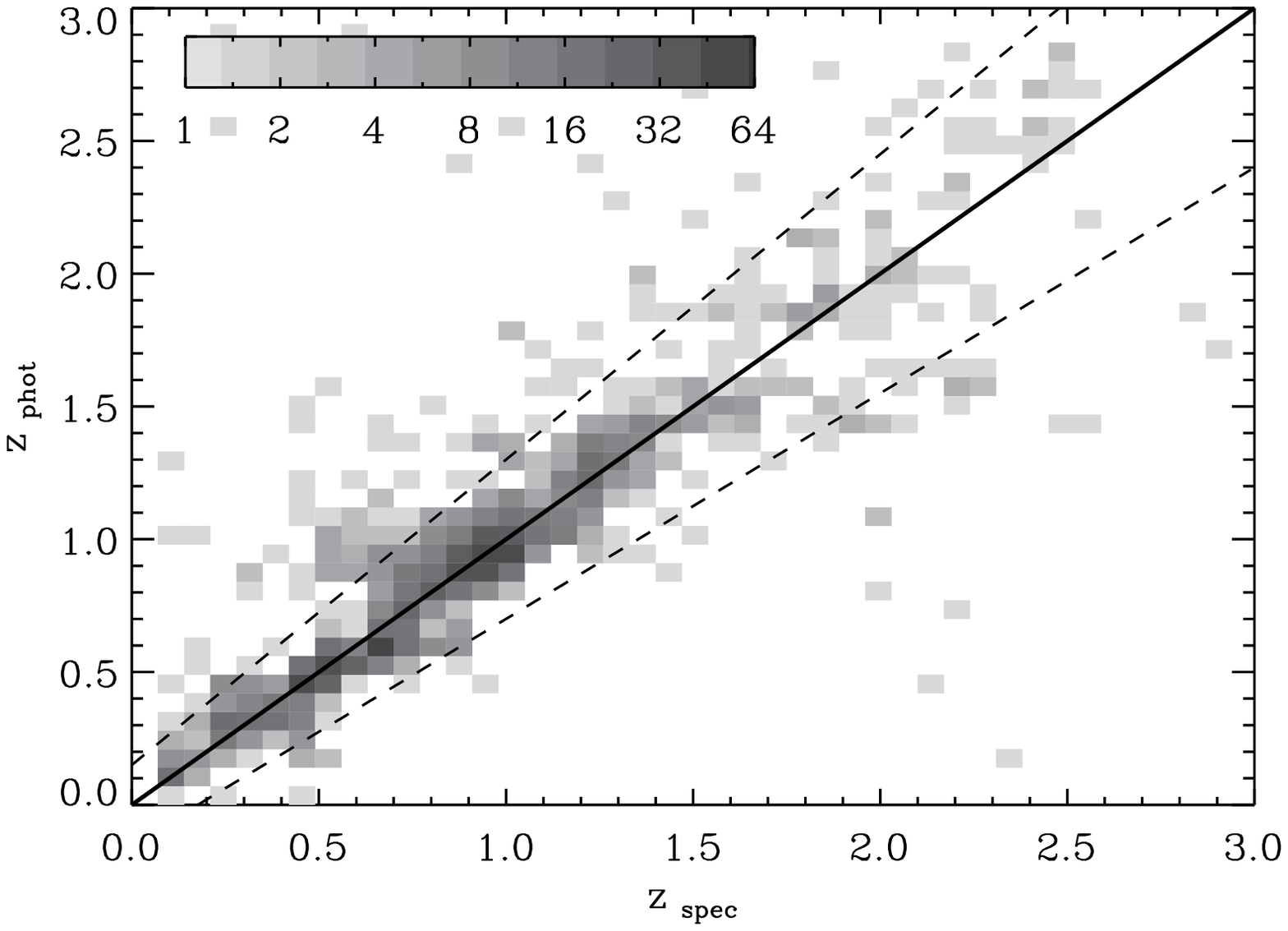}
	\includegraphics[width=8.cm]{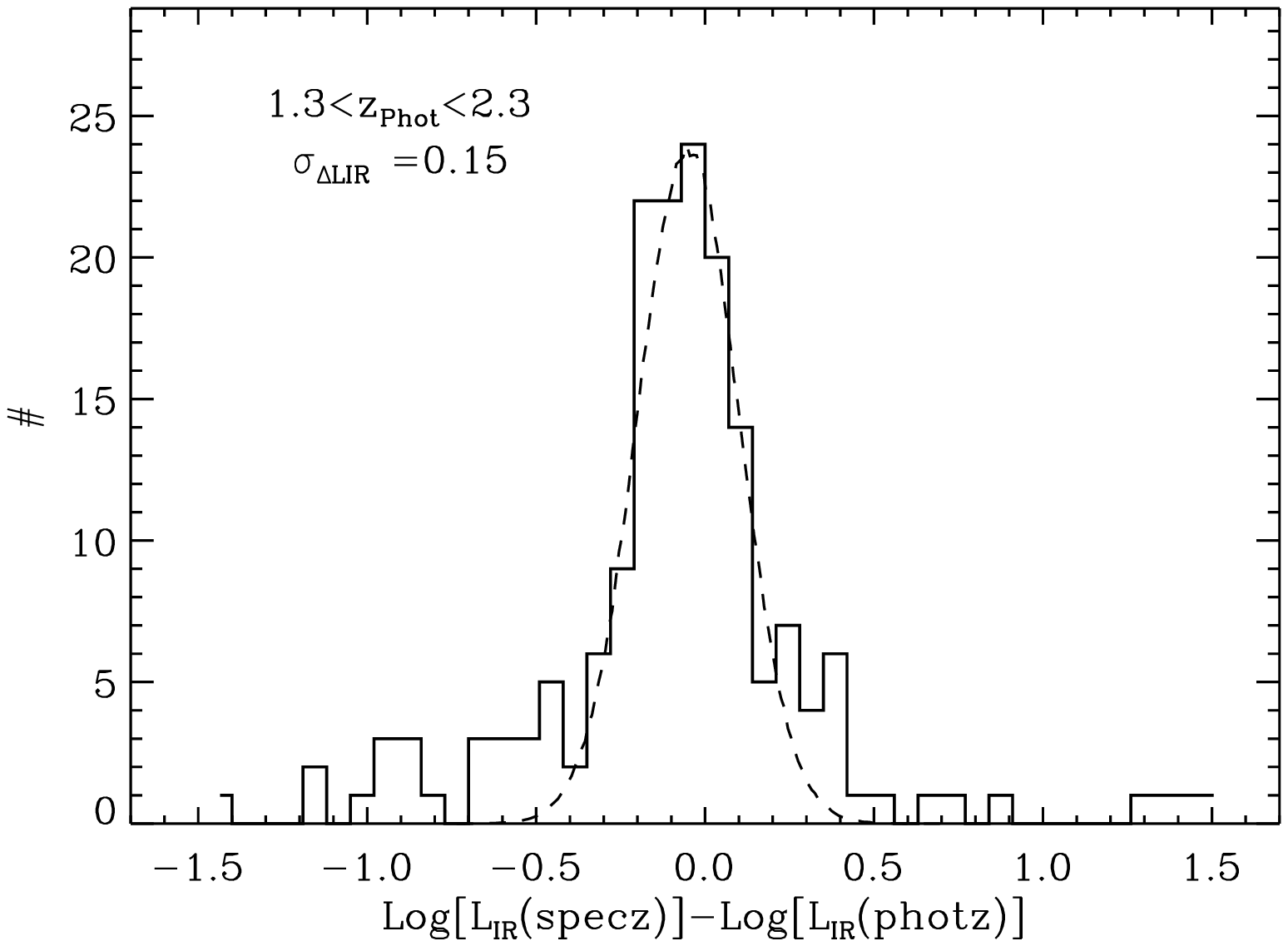}
	\caption{\label{fig:data}\textbf{\textit{(Left)}} Comparison between the photometric and spectroscopic redshifts of our 24 $\mu$m-selected catalog.
	This comparison is made using 1670 galaxies which have both kinds of redshifts.
	Dashed lines represent the relative errors found in the redshift range of our study (i.e., $\sigma_{\Delta z/(1+z)}=0.14$ for $1.3<z<$2.3).
	\textbf{\textit{(Right)}} Uncertainty in determining L$_{\rm{IR}}$ from 24~$\mu$m fluxes densities and the \citet{chary_2001} library due to error in photometric redshift estimates (\textit{black histogram}).
	The dashed line is a Gaussian fit to this uncertainty distribution with $\sigma=0.15$.
	Note that the uncertainty distribution is not Gaussian.
	}
\end{figure*}
\section{Data}
\subsection{Infrared Imaging\label{subsec:infrared data}}
\indent{
The 24 $\mu$m imaging of the GOODS-N ($12^{h}36^{m},\,+62^{\circ}14\arcmin$) and GOODS-S ($3^{h}32^{m},\,-27^{\circ}48\arcmin$) fields were obtained as part of the GOODS Legacy program (PI: M. Dickinson).
The 70~$\mu$m data in both GOODS fields were obtained by \textit{Spitzer} GO programs GO-3325 and GO-20147 (PI: Frayer).
In the north they cover a region of roughly $10\arcmin\times16\arcmin$, while in the south they cover a somewhat smaller fraction of the GOODS 24 $\mu$m area, roughly $10\arcmin\times10\arcmin$.
The Frayer data have been combined with additional 70 $\mu$m observations covering a wider area from the Far-Infrared Deep Extragalactic Legacy (FIDEL) program (PI: Dickinson), as well as shallower data from \textit{Spitzer} Guaranteed Time Observer (GTO) programs (PI: Rieke).
While our catalog is complete to the observational limits, here we restrict our analysis to the deepest regions at 70 $\mu$m covered by the Frayer+FIDEL+GTO data, covering areas of 194~arcmin$^2$ and 89~arcmin$^2$ in GOODS-N and -S, respectively.
\\}
\indent{
At the resolution of \textit{Spitzer} at 24 and 70 $\mu$m, all sources in our fields are point sources (i.e FWHM$\thicksim5.9\arcsec$ and 18\arcsec at 24 $\mu$m and 70 $\mu$m respectively).
Flux densities at these wavelengths are hence estimated using a PSF fitting technique based on the knowledge of the expected positions of the sources \citep{magnelli_2009}.
For the 24 $\mu$m data, we use the position of the IRAC 3.6 $\mu$m sources as priors.
This choice is motivated by the fact that the IRAC 3.6~$\mu$m data are 30 times deeper than our current 24~$\mu$m observations and that the typical $S_{24\,\mu m}/S_{3.6\,\mu\rm{m}}$ ratio spans the range [2-20] \citep{chary_2004}.
Hence we can assume that all 24 $\mu$m sources have an IRAC 3.6 $\mu$m counterpart.
For the 70~$\mu$m data we use as prior the IRAC positions of our flux limited sample of 24 $\mu$m sources.
At this wavelength our choice is straightforward since the typical $S_{70\,\mu m}/S_{24\,\mu m}$ ratio spans the range $[2-100]$ \citep{papovich_2007} and the 24 $\mu$m observations are about 100 times deeper than our current 70 $\mu$m observations.
These assumptions are tested by visual inspection of residual images which would reveal any 24~$\mu$m or 70 $\mu$m sources missed due to the lack of priors.
We find no such sources.
\\}
\indent{
For GOODS-N, we use the IRAC catalog generated from the publicly available GOODS Legacy data ($19437$ objects detected at 3.6 $\mu$m with a $50\%$ completeness limit of $0.5\,\mu$Jy).
The GOODS-S IRAC data have been incorporated into the SIMPLE IRAC Legacy Survey observations covering the wider Extended Chandra Deep Field South, also observed with MIPS as part of the FIDEL Legacy program. 
We use the SIMPLE IRAC catalogs \citep[][$61233$ objects detected at 3.6~$\mu$m with a 50\% completeness limit of 1.5~$\mu$Jy]{damen_2011} as priors for the MIPS source extraction over the whole FIDEL area, although here we concentrate only on the deep $10\arcmin\times10\arcmin$ GOODS-S region.
\\}
\indent{
Using Monte carlo simulations we are able to estimate the quality of our 24 and 70~$\mu$m catalogs \citep[see][]{magnelli_2009}.
We find that, in both fields, the 24 and 70 $\mu$m observations reach an $80\%$ completeness limit of 30 $\mu$Jy and $2.5$~mJy respectively.
In the GOODS-N and S fields we detect 2151 and 870 sources respectively with $S_{24}>30 \,\mu$Jy, and 119 and 50 sources with $S_{70}>2.5$~mJy.
Tables \ref{tab:sources 1},  \ref{tab:sources 2}, \ref{tab:sources 3} and \ref{tab:sources 4} of the \textit{online material} give excerpt of the complete GOODS-N/S 24 $\mu$m and 70 $\mu$m catalogs that are now available on CDS\footnote{http://cdsweb.u-strasbg.fr/cgi-bin/qcat?J/A+A/$<$volume$>$/$<$page$>$}.
These catalogs extend below the $80\%$ completeness limit, and covers the full extent (approximately $10\arcmin\times16\arcmin$) of the GOODS-S region, not only the smaller $10\arcmin\times10\arcmin$ region with the deepest 70 $\mu$m imaging that is used for the analysis in this paper.
\\ \\}
\indent{
Since calibration factors taken to generate the final 24 and 70 $\mu$m mosaics are derived from stars, color corrections (at most $\thicksim 10\%$) have to be apply to all our fluxes.
These color corrections being highly dependent on the redshift and the intrinsic SED of the sources, we decided to introduce these color corrections directly in the \textit{k}-correction used to estimate the LF since both quantities are taken into account in this computation.
\\ \\}
\subsection{Redshifts\label{subsec:red}}
\indent{
In this study we use spectroscopic redshifts coming from a combination of various studies  \citep[][Kurk et al. in prep for GMASS redshifts and finally Stern et al. in prep]{cohen_2000,wirth_2004,cowie_2004,lefevre_2004,mignoli_2005,vanzella_2006,reddy_2006b,barger_2008,cimatti_2008}.
Photometric redshifts are computed using Z-PEG \citep{leborgne_2002} and all optical and near infrared data currently publicly available.
In GOODS-N, optical observations in the \textit{BVIz} passbands were obtained with the \textit{Advanced Camera for Surveys} (ACS) onboard the \textit{Hubble Space telescope} (HST) as part of the GOODS ACS Treasury program (M. Giavalisco and the GOODS Team, 2010, in preparation) while near infrared observations in the \textit{JK} passbands were taken from the KPNO 4m FLAMINGOS catalog.
In GOODS-S optical and near infrared observations were taken from the GOODS MUSIC catalog \citep[][\textit{UBVIzJHK}]{grazian_2006,santini_2009}.
\\}
\begin{table*}
\caption{\label{tab: sample} Redshift catalog properties}
\center{
\begin{tabular}{ccccccccc}
\hline \hline
Field & Area & Nb  sources & X-ray AGN & No X-ray AGN & \# spec-$z$$^{\,\mathrm{a}}$ & \# phot-$z$$^{\,\mathrm{a,b}}$ & \# spec-$z$ and/or phot-$z$$^{\,\mathrm{a}}$ \\
 & \tiny {($\rm{arcmin^{2}}$)}  & \tiny {$\rm{24 \mu m/ 70 \mu m}$} &  \tiny{$\rm{24 \mu m/70 \mu m}$}  &  \tiny{$\rm{24 \mu m/70\mu m}$}  & \tiny{$\rm{24 \mu m/70 \mu m}$}  & \tiny{$\rm{24 \mu m/70 \mu m}$}  & \tiny{$\rm{24 \mu m/70 \mu m}$} \\
\hline
GOODS-S & 89 & $870/44$ & $64/7$ & $806/37$ & $371/32$ & $378/5$ & $749/37$\\
 & &  & \tiny{$7\%/16\%$} & \tiny{$93\%/84\%$} & \tiny{$46\%/86\%$} & \tiny{$46\%/14\%$} & \tiny{$92\%/100\%$}\\
\\
GOODS-N & 194 & $2151/119$ & $134/12$ & $2017/107$ & $747/72$ & $1148/29$ & $1895/101$\\
&  &  & \tiny{$6\%/10\%$} & \tiny{$94\%/90\%$} & \tiny{$37\%/67\%$} & \tiny{$57\%/27\%$} & \tiny{$94\%/94\%$}\\
\\
\hline
\end{tabular}
}
\begin{list}{}{}
\item[$^{\mathrm{a}}$] percentages noted in these columns refer to the number of no X-ray AGN
\item[$^{\mathrm{b}}$] number of sources which have a photometric redshift but no spectroscopic redshift
\end{list}
\end{table*}
\subsection{Removing AGNs}
\indent{
To identify and remove X-ray AGN we use deep \textit{Chandra} X-Ray observations, i.e., the 1~Ms maps for GOODS-S and the 2 Ms maps  for GOODS-N \citep{alexander_2003}.
AGNs are identified as galaxies with either $L_{X}\, [0.5\,-\,8.0\ \rm{keV}]\,\ge\,3\,\times\,10^{42}\ \rm{erg\, s^{-1}}$ or a hardness ratio greater than 0.8 (ratio of the counts in the $2 - 8$ keV to $0.5-2$ keV passbands) \citep{bauer_2004}.
Even if it is well-known that AGN do also harbor star formation, we do not subtract the AGN contribution to the infrared light of those galaxies since such subtraction would be highly speculative at the present level of our knowledge.
Instead we conservatively decide to remove all those galaxies from our sample.
\\}
\subsection{The final infrared galaxy sample}
\indent{
To construct our final infrared sample we first cross-match the 24~$\mu$m catalog with the X-ray observations.
We find that $\thicksim 6\%$($13\%$) of the 24(70) $\mu$m sources contain an X-ray AGN.
All those sources are excluded from our final infrared sample.
Remaining sources are then matched with our spectroscopic and photometric redshift catalogs, using a matching radius of $1.5\,\arcsec$ (i.e. $\thicksim FWHM$ of the IRAC $3.6\,\mu$m observations).
In case of multiple associations we select the closest optical counterparts.
In GOODS-N (GOODS-S) $94\%$ ($92\%$) of the 24~$\mu$m sources brighter than 30 $\mu$Jy have a spectroscopic and/or a photometric redshift and $80\%$ (76\%) of these sources have been detected in the near infrared.
$46\%$ and $37\%$ of our 24~$\mu$m sources have a spectroscopic redshift in GOODS-S and -N respectively.
\\}
\indent{
All the different steps described previously are listed in the Table \ref{tab: sample}.
These steps yield to a final infrared galaxy sample containing 2644 and 138 sources detected at 24 $\mu$m and 70 $\mu$m respectively.
In our redshift range of interest (i.e. $1.3<z<2.3$) our infrared galaxy sample contains 706 sources detected at 24~$\mu$m and only 8 sources detected at 70 $\mu$m.
In this redshift range, the fraction of sources with a spectroscopic redshift is $\thicksim20\%$.
\\ \\}
\begin{figure}
	\includegraphics[width=9.5cm]{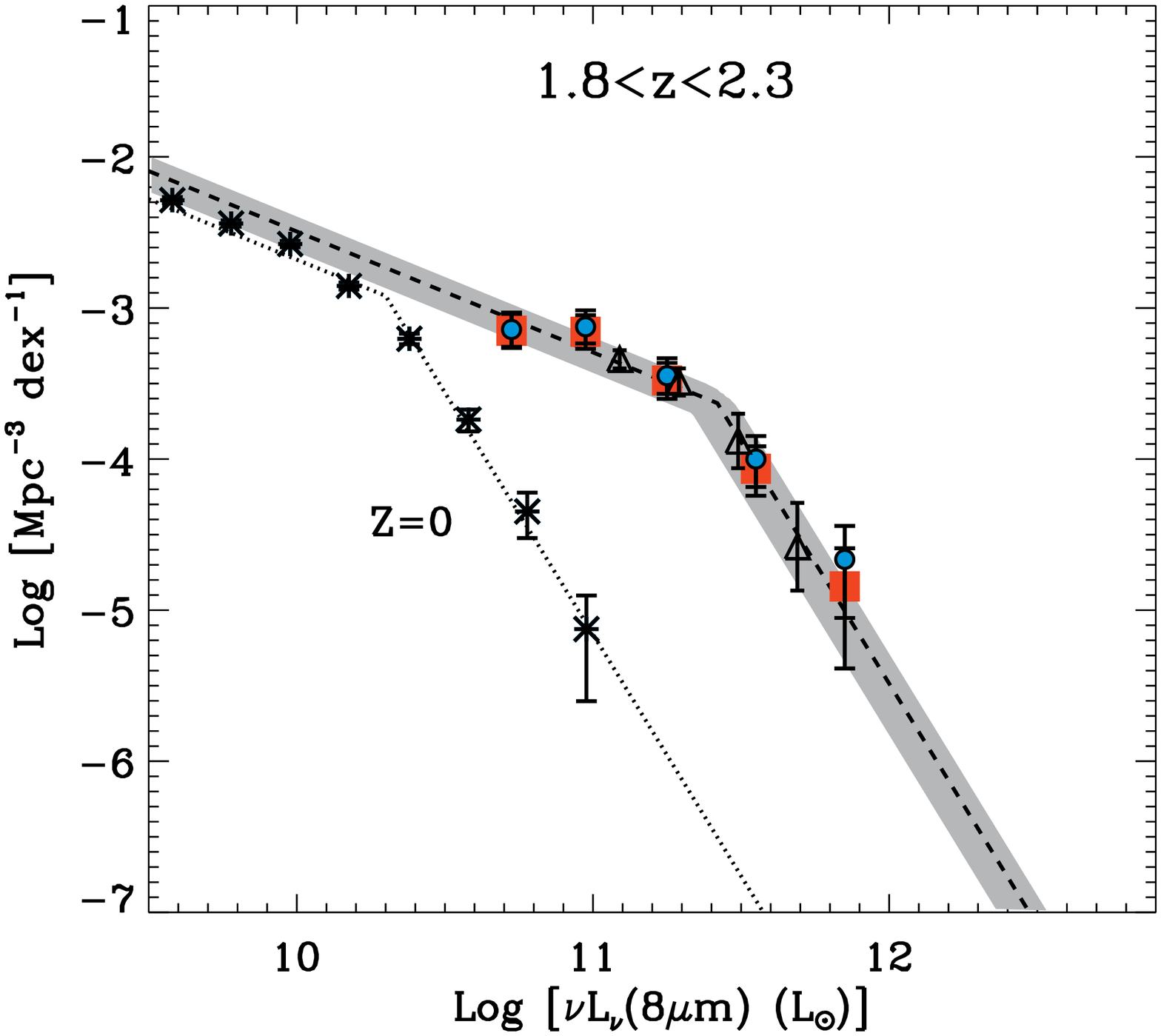}
	\caption{\label{fig:LF 8 seul}Rest-frame 8 $\mu$m LF estimated at $z\thicksim2$ using the $1/V_{max}$ method.
	Light blue circles and red squares represent the rest-frame 8 $\mu$m LF derived from our 24 $\mu$m sample with and without X-ray AGNs respectively.
	Empty triangles represent the rest-frame 8 $\mu$m LF derived by \citet{caputi_2007} at $z\thicksim2$.
	Asterisks show the local reference taken from \citet{huang_2007} and the dotted line represents the best-fit to these data points with a double power law function with fixed slopes (i.e., $\phi \propto L^{-0.8}$ for $L<L_{knee}$ and  $\phi \propto L^{-3.2}$ for $L>L_{knee}$).
	The dashed line represents the best fit of the rest-frame 8 $\mu$m LF at $z\thicksim2$ assuming that the shape of the rest-frame 8 $\mu$m LF remains the same since $z\thicksim0$.
	The dark shaded area span all the solutions obtained with the $\chi^{2}$ minimization method and compatible, within 1 $\sigma$, with our data points.
		}
\end{figure}
\indent{
In Figure \ref{fig:data} (\textit{left}), we compare spectroscopic and photometric redshifts of 1670 sources detected at 24 $\mu$m and with both kinds of redshifts (i.e. spectroscopic and photometric redshifts).
In the $1.3<z<2.3$ redshift range, accuracy of the photometric redshifts is $\sigma_{\Delta z/(1+z)}=0.14$ and $\Delta z/(1+z)$ has a median value of $-0.002$.
These redshift uncertainties result in infrared luminosity uncertainties when converting 24~$\mu$m flux densities into $L_{IR}$ using the \citet{chary_2001} library (see Figure \ref{fig:data}, \textit{right}).
Since the redshift uncertainties are not Gaussian, the infrared luminosity uncertainties are also not Gaussian: wings of the real distribution extend further away than in a Gaussian distribution.
As a result, to study the real impact of redshift uncertainties on the inferred infrared LF, one needs to introduce the real redshift distribution into Monte-Carlo simulations instead of using a redshift distribution with a Gaussian statistic.
Such Monte-Carlo simulations have been done and are discussed in section \ref{subsec:monte}.
\\ \\}
\indent{
To illustrate the impact on the inferred LF of the subtraction of the X-ray AGNs, we compute at $z\thicksim2$ the rest-frame 8 $\mu$m LF with and without X-ray AGNs.
The choice of this particular redshift and wavelength is motivated by the fact that at $z\thicksim2$ the 24 $\mu$m observations correspond to the rest-frame 7.8\,$\mu$m.
Therefore the extrapolation that needs to be applied to compute the 8 $\mu$m luminosity is negligible, nearly independent of the SED library used, as well as independent of the nature of the source (i.e., AGN or star-forming galaxy).
In Figure~\ref{fig:LF 8 seul} we present the rest-frame 8~$\mu$m LF derived from the $1/V_{max}$ analysis (see section \ref{subsec:monte} for a precise description of this method) and using the \citet{chary_2001} library.
\\}
\indent{
We note that the LF derived with and without the AGN are in total agreement except for the brightest luminosity bin.
Even for this last luminosity bin the difference between these two LFs is relatively small and is of the order of $\thicksim0.2$ dex.
As a result, we can conclude that our particular choice for handling AGN will not affect our final results much and will not be able to explain the large discrepancies that will arise in later sections when we compare our LFs with others that have appeared in the literature.
\\}
 \indent{
Our rest-frame 8 $\mu$m LF is in excellent agreement with the one inferred by \citet{caputi_2007}, confirming the consistency of our 24 $\mu$m sample.
We note that the LF derived in our study extends to fainter luminosities since we are using a 24~$\mu$m catalog that is $\sim 3$ times deeper.
\\}
\section{The Total Infrared Bolometric Correction\label{subsec:LIR}}
\begin{figure*}
\centering
	\includegraphics[width=16cm]{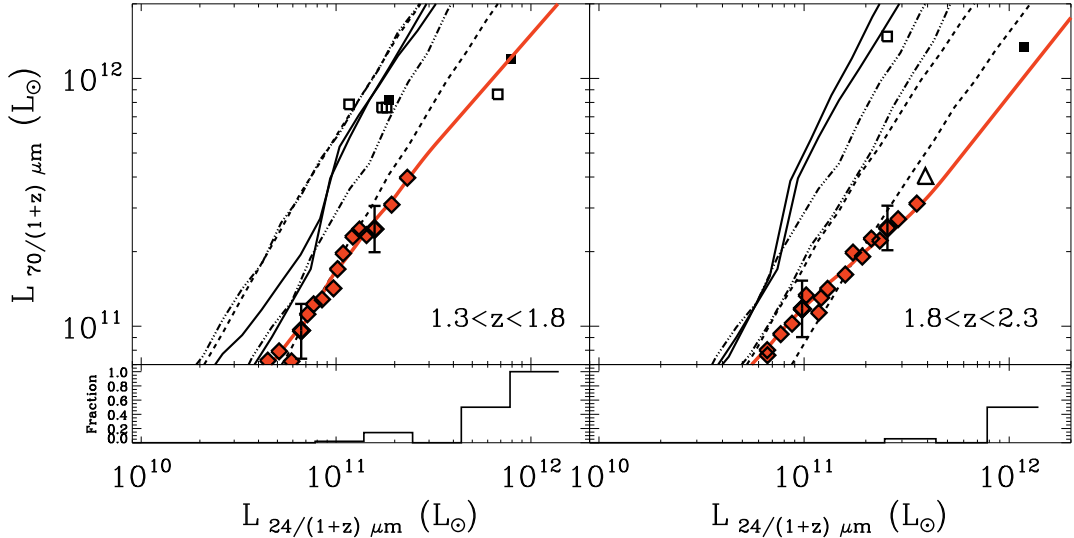}
	\caption{\label{fig:stacking}The 70 vs. 24 $\mu$m correlations as revealed by the observations and our stacking analysis in the two redshift bins considered in this study.
	The empty and filled squares represent the 70 vs. 24 $\mu$m correlations observed for sources individually detected at 70~$\mu$m with photometric or spectroscopic redshift respectively.
	The red diamonds show the results obtained using our stacking analysis (see text).
	For clarity the error bars of our stacking analysis are shown only for two points.
	These error bars are computed using a standard bootstrap analysis and an estimate of background fluctuation using a stacking analysis at random positions \citep[for more detail see][]{magnelli_2009}.
	The empty triangle shows the median correlation found in the sample of \citet{murphy_2009}.
	The thin solid lines, the dashed lines and the triple-dots-dash lines represent the expected correlations for the CE01, the LDP and the DH libraries, respectively, at the lowest and the highest redshift of each redshift bin.
	The red solid line represents the inferred 24/70~$\mu$m correlation derived using a smooth linear interpolation between red diamonds and extended at high luminosities using 24 $\mu$m sources individually detected at 70 $\mu$m.
	At the bottom of each plot we present the fraction of 24 $\mu$m sources that are individually detected at 70 $\mu$m as a function of the 24/(1+z)~$\mu$m luminosity.
	}
\end{figure*}
\begin{figure}
	\includegraphics[width=9.cm]{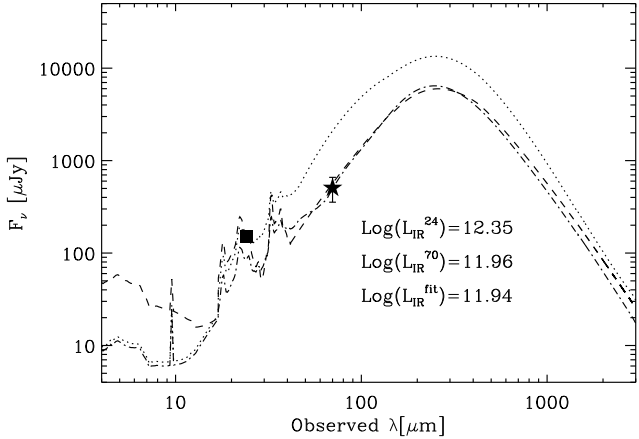}
	\caption{\label{fig:agn pah} Different infrared bolometric corrections applied to a $z\thicksim1.8$ galaxy.
	Black square represents the observed 24 $\mu$m flux density while the black star represents the 70 $\mu$m flux density predicted using the $L_{24\,\mu \rm{m/(1+z)}}-L_{70\,\mu \rm{m/(1+z)}}$ correlation.
	Dotted line and the dashed-dot line represent the unscaled CE01 templates corresponding respectively to the observed 24~$\mu$m and the predicted 70 $\mu$m flux densities.
	The dashed line represents the scaled CE01 template which best fit the 24 and 70~$\mu$m flux densities of this galaxy.
	$L_{IR}^{24}$, $L_{IR}^{24}$ and $L_{IR}^{fit}$ give the infrared luminosity derived by integrating the dotted line, the dashed-dot line and the dashed line respectively.
	}
\end{figure}

\indent{
In the local Universe tight correlations have been found between monochromatic (eg, $L_{15\,\mu \rm{m}}$, $L_{24\,\mu \rm{m}}$ and $L_{70\,\mu \rm{m}}$ ... etc) and total infrared luminosities ($L_{\rm{IR}}=L$\,[8-1000 $\mu$m]) of galaxies \citep{chary_2001}.
Based on these correlations SED libraries have been developed and extensively used to estimate the total infrared luminosity of galaxies \citep[e.g.,][hereafter CE01, LDP, and DH\footnote{We note that the Dale \& Helou library is originally parametrized using the IRAS far-infrared colors (i.e., $R60/100$). Nevertheless the one used here has been parametrized a posteriori with $L_{IR}$ using the local $R60/100$ vs $L_{IR}$ correlation \citep{soifer_1991}.} respectively]{chary_2001,lagache_2003,dale_2002}.
However, it is not clear that these local templates are suitable to describe the spectral properties of distant galaxies  \citep{papovich_2007,daddi_2007a,leborgne_2009}.
To study this issue, we decided to characterize the observed 24~$\mu$m vs 70~$\mu$m correlation and compare it to the predictions of standard SED libraries.
Since at high redshift (i.e. $z>1.3$) most of the 24~$\mu$m sources are undetected at 70~$\mu$m, the characterization of the 24~$\mu$m vs 70~$\mu$m correlation could only rely on mean 70 $\mu$m properties obtained through a stacking analysis.
This method, which has been extensively used in the last few years, gives reliable estimates of the typical 70 $\mu$m flux density of a given galaxy population even below the detection limit of current 70 $\mu$m observations \citep[e.g.][]{dole_2006,papovich_2007,magnelli_2009}.
\\}
\indent{
We first divided our 24 $\mu$m sample into two redshift bins, of $1.3<z<1.8$ and $1.8<z<2.3$.
Then, inside these redshift bins, we separated these 24 $\mu$m sources per luminosity bins of 0.5 dex.
For each 24 $\mu$m luminosity bin we stacked on the residual 70\,$\mu$m image all sources with no 70~$\mu$m detection.
The photometry of the stacked image was then measured using an aperture radius of 16\arcsec, a background from within annuli of 40\arcsec and 60\arcsec and an aperture correction factor of 1.705 (as discussed in the \textit{Spitzer} observer's manual).
Finally, the mean 70\,$\mu$m flux density ($F^{70\,\mu m}_{bin}$) for a given 24~$\mu$m luminosity bin was computed following Equation \ref{eq:mean stack} :
\begin{center}
\begin{equation}\label{eq:mean stack} 
F^{70\,\mu m}_{bin}=\frac{m\times F^{70\,\mu m}_{stack}+\sum_{i=1}^{n}F^{70\,\mu m}_{i}}{n+m}
\end{equation}
\end{center}
where $F^{70\,\mu m}_{stack}$ is the stacked 70 $\mu$m flux density of all 24~$\mu$m sources within this luminosity bin and undetected at 70~$\mu$m (sample which contains $m$ sources);
$F^{70\,\mu m}_{i}$ is the 70~$\mu$m flux density of the $i$th 24~$\mu$m sources within this luminosity bin and detected at 70~$\mu$m (sample which contains $n$ sources).
This procedure was performed using sliding 24~$\mu$m luminosity bins with steps of 0.1 dex.
While such small sliding steps introduce correlations between our staking results, it avoids problems one might introduce by arbitrarily choosing some particular luminosity bins.
Moreover, we note that since our staking analysis probes a dynamic range of $\thicksim1.5$ dex along the $L_{24\,\mu \rm{m/(1+z)}}$ axis, there are always three independent measurements to characterize the typical $L_{24\,\mu \rm{m/(1+z)}}$-$L_{70\,\mu \rm{m/(1+z)}}$ correlation.
\\ \\}
\indent{
Results of this stacking analysis are shown with filled red diamonds in Figure \ref{fig:stacking}.
In both redshift bins, we find that none of the usual SED libraries can reproduce the observed correlation between $L_{24\,\mu \rm{m/(1+z)}}$ and $L_{70\,\mu \rm{m/(1+z)}}$.
At high $24\,\mu \rm{m/(1+z)}$ luminosities, standard SED libraries predict a higher $L_{70\,\mu \rm{m/(1+z)}}$ to $L_{24\,\mu \rm{m/(1+z)}}$ ratio than is actually observed.
These low $L_{70\,\mu \rm{m/(1+z)}}$ to $L_{24\,\mu \rm{m/(1+z)}}$ ratio could be reproduced by standard SED libraries but would correspond to SED templates with very low intrinsic $24\,\mu \rm{m/(1+z)}$ luminosities.
In other words, the actual normalization of standard SED libraries which predict the increase of the $L_{70\,\mu \rm{m/(1+z)}}$ to $L_{24\,\mu \rm{m/(1+z)}}$ ratio with increasing $24\,\mu \rm{m/(1+z)}$ luminosity is wrong.
Instead, we observed that at high redshift the $L_{70\,\mu \rm{m/(1+z)}}$ to $L_{24\,\mu \rm{m/(1+z)}}$ ratio does not strongly depend on the $24\,\mu \rm{m/(1+z)}$ luminosity and that sources with high $24\,\mu \rm{m/(1+z)}$ luminosity (or equivalently high infrared luminosity) have a $L_{70\,\mu \rm{m/(1+z)}}$ to $L_{24\,\mu \rm{m/(1+z)}}$ ratio typical of sources in the local universe with low $24\,\mu \rm{m/(1+z)}$ luminosity.
\\}
\indent{
We note that among the few 24 $\mu$m sources detected at 70 $\mu$m (\textit{open} and \textit{filled squares} in figure \ref{fig:stacking}) those with high $24\,\mu \rm{m/(1+z)}$ luminosities (i.e. $L_{24\,\mu \rm{m/(1+z)}}>8\times10^{11}\,$L$_{\odot}$) confirm the discrepancy inferred using the stacking analysis.
On the contrary, those sources with low $24\,\mu \rm{m/(1+z)}$ luminosities are significantly closer to the predictions of standard SED libraries.
This behavior is of course driven by selection effects and those sources only represent the high-end tail of the dispersion of the $L_{24\,\mu \rm{m/(1+z)}}$ vs $L_{70\,\mu \rm{m/(1+z)}}$ correlation inferred through stacking.
\\}
\indent{
We note that the discrepancies that we find between the observed $L_{24\,\mu \rm{m/(1+z)}}$ vs $L_{70\,\mu \rm{m/(1+z)}}$ correlation and predictions from standard SED libraries is quite different than what we found previously for galaxies at $z<1.3$ using a similar stacking analysis \citep{magnelli_2009}.
There, the CE01 models provided a reasonably good fit to the observed $L_{24\,\mu \rm{m/(1+z)}}$ vs $L_{70\,\mu \rm{m/(1+z)}}$ correlation.
\\ \\}
\indent{
Our findings indicate that the total infrared luminosity of a high-redshift galaxy cannot be inferred simply using its 24 $\mu$m flux density and any of the three standard SED libraries considered here.
We note that this analysis is fully consistent with the results presented by \citet[][see their figure 8]{daddi_2007a} and first results obtained using \textit{Herschel} data \citep{nordon_2010}.
At $z \sim 2$ and at $L_{24\,\mu \rm{m/(1+z)}}=8\times10^{10}$ and $3.5\times10^{11}$~L$_{\odot}$, \citet{daddi_2007a} find that  
$L_{70\mu{\rm m}/(1+z)}$ would be overestimated by a factor $\thicksim2$ and $\thicksim10$ using the CE01 library respectively, while we find a factor 2.3 and 12 respectively.
\\}
\indent{
In \citet{papovich_2007} and in \citet{murphy_2009} similar discrepancies were found and explained in part as the result of an increase in the PAH emission at any given infrared luminosity in high-redshift galaxies.
In both studies, the infrared luminosity of galaxies were then derived using colder SED templates (i.e., like local galaxies with lower infrared luminosities) which fit the observed correlation.
\\}
\indent{
An alternative explanation for the discrepancies between SED libraries and the observed correlations can be the presence of an obscured AGN.
Indeed, as suggested by \citet{daddi_2007}, the observed 24 $\mu$m flux density of a galaxy located at $z\thicksim2$ might be dominated by the hot dust continuum from an obscured AGN and hence it might not be a robust SFR indicator as inferred from its disagreement with radio stacking and extinction corrected UV estimates.
In the same study, \citet{daddi_2007a} show that in this redshift range ($1.5<z<2.5$) the SFRs derived from radio, extinction corrected UV and 70~$\mu$m flux density agree well.
This suggests that unlike 24~$\mu$m, the 70~$\mu$m passband is a good SFR indicator even in high-redshift galaxies.
Thus, we will use the observed $L_{24\,\mu \rm{m/(1+z)}}-L_{70\,\mu \rm{m/(1+z)}}$ correlation to infer $L_{70\,\mu \rm{m/(1+z)}}$ for each 24 $\mu$m source and use this in turn to derive its total infrared luminosity and star-formation rate.
\\ \\}
\indent{
Based on these two explanations we decided to derive the total infrared luminosity of each galaxy using two different methods.
\textit{(i)} For each 24 $\mu$m source we deduce its 70 $\mu$m flux density using the $L_{24\,\mu \rm{m/(1+z)}}-L_{70\,\mu \rm{m/(1+z)}}$ correlation.
Then we choose the CE01 template whose redshifted color best matches the derived 24 to 70~$\mu$m flux ratio and renormalized it to match the 24 $\mu$m flux density of the source (i.e., ignoring the intrinsic luminosity normalization of the CE01 library; \textit{Dashed line} of Fig. \ref{fig:agn pah}).
The total infrared luminosity of this galaxy (hereafter $L_{IR}^{fit}$) is estimated by integrating the SED curve of this template.
In the following, we will assume that the uncertainties on the infrared luminosity estimated using this method is of order 0.2 dex as measured by \citet{murphy_2009}.
We note that the $L_{70\,\mu \rm{m/(1+z)}}/L_{24\,\mu \rm{m/(1+z)}}$ ratio is nearly constant over the whole $L_{24\,\mu \rm{m/(1+z)}}$ luminosity range and corresponds to that expected for a CE01 template with an intrinsic $Log(L_{IR}\,[L_{\odot}]\,)\thicksim10.8$.
Such SED templates exhibit strong PAH features.
\textit{(ii)} For each 24 $\mu$m source, a 70 $\mu$m flux density is derived using the $L_{24\,\mu \rm{m/(1+z)}}-L_{70\,\mu \rm{m/(1+z)}}$ correlation.
$L_{IR}$  (hereafter $L_{IR}^{70}$) is then simply estimated using the CE01 template (i.e., keeping the intrinsic luminosity normalization) matching the derived value of $L_{70\mu{\mathrm m}/(1+z)}$.
\\ \\}
\indent{
Figure \ref{fig:agn pah} illustrates these two different bolometric corrections for a galaxy situated at $z=1.8$.
We note that fitting the 24 and 70 $\mu$m measurements together, allowing the renormalization of the CE01 SED templates, or using the luminosity-normalized CE01 library to fit the 70 $\mu$m alone give nearly the same results (for all our 24 $\mu$m sample we find $<Log(L_{IR}^{fit}/L_{IR}^{70})>\thicksim0.04$ and $\sigma[Log(L_{IR}^{fit}/L_{IR}^{70})]\thicksim0.05$).
Indeed, both techniques involve the use of SED templates with lower intrinsic infrared luminosity than from the 24 $\mu$m alone.
We also note that using the 24 $\mu$m flux density alone and the luminosity-normalized CE01 library we would have overestimated the total infrared luminosity of this galaxy by $\thicksim0.4$ dex. 
\\}
\section{The Infrared Luminosity Function}
\subsection{Methodology \label{subsec:monte}}
\indent{
The infrared LFs are derived using the standard $1/V_{max}$ method \citep{schmidt_1968}.
The comoving volume associated with any source of a given luminosity is defined as $V_{max}=V_{zmax}-V_{zmin}$ where $zmin$ is the lower limit of the redshift bin, and $zmax$ is the maximum redshift at which the object could be seen given the flux density limit of the sample, with a maximum value corresponding to the upper limit of the redshift bin.
For each luminosity bin, the LF is then given by
\\}
\begin{equation}\label{eq:sfr lir}  
	\phi = \frac{1}{\Delta L}\sum_{i}\frac{1}{V_{max,i}\times w_{i}}
\end{equation}
\indent{
where $V_{max}$ is the comoving volume over which the $i$th galaxy could be observed, $\Delta L$ is the size of the luminosity bin, and $w_{i}$ is the completeness correction factor of the $i$th galaxy.
$w_{i}$ equals 1 for sources brighter than $S_{24\mu m}\thicksim 100\,\mu$Jy and decreases at fainter flux densities due to the incompleteness of the 24 $\mu$m catalog.
These completeness correction factors are robustly determined using Monte-carlo simulations \citep{magnelli_2009} and reach at most a value of 0.8. 
None of the conclusions presented here strongly rely on this correction.
\\ \\}
\indent{
Uncertainties in the infrared LF values depend on photometric redshift uncertainties (see Figure \ref{fig:data}).
In particular, catastrophic redshift errors (i.e. $\Delta z/(1+z)>0.15$), which can shift a low redshift galaxy to higher redshift and {\it vice versa}, can modify the number density of LIRGs and ULIRGs in a given redshift bin.
To estimate the effect of these catastrophic redshift errors on the derived infrared LF, one needs to have a complete census on the population of infrared galaxies at all redshifts.
To simulate this, we first generate a reference catalog (i.e., an ideal sample with no redshift uncertainties), then from this reference catalog we generate 1000 mock catalogs with realistic redshift uncertainties and finally we compare the infrared LF inferred from the reference catalog with the infrared LF inferred from the 1000 mock catalogs.
The key point of these simulations is to introduce redshift uncertainties which accurately reproduce the observed distribution of spectroscopic versus photometric redshifts (see Figure \ref{fig:data}) instead of using a standard Gaussian distribution which would not be realistic (see discussion in section \ref{subsec:red}).
\\ \\}
\begin{figure*}
\centering
	\includegraphics[width=17.5cm]{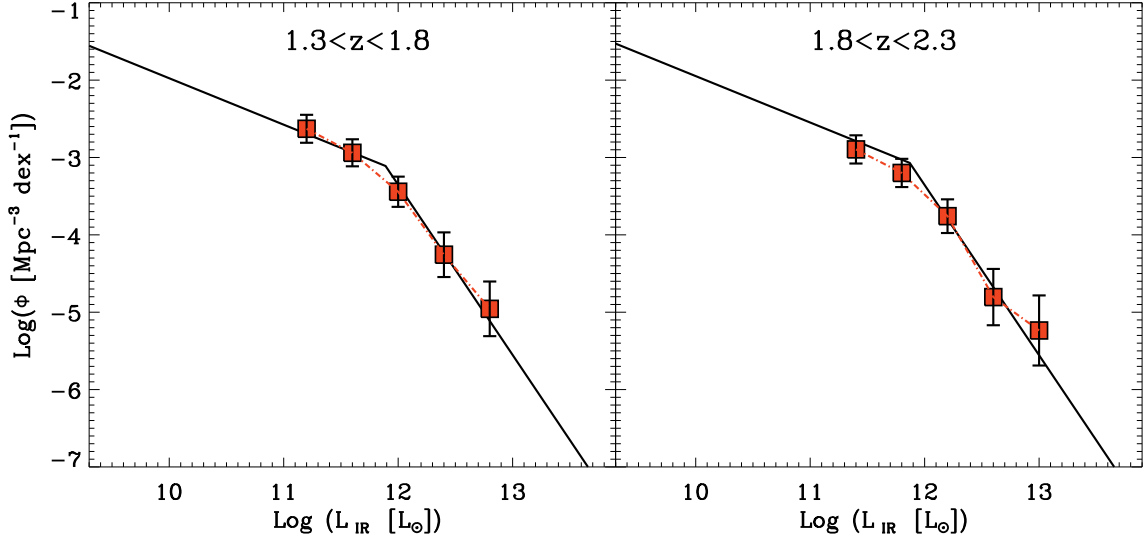}
	\caption{\label{fig:monte} Results of our Monte Carlo simulations.
	Black lines represent the infrared LF that one would have inferred using an ideal sample with no redshift or bolometric correction uncertainties.
	Red squares show the mean infrared LF inferred using our 1000 mock catalogs.
	Error bars correspond to the dispersion observed in our 1000 Monte Carlo simulations.
	}
\end{figure*}
\begin{figure*}
\centering
	\includegraphics[width=17.5cm]{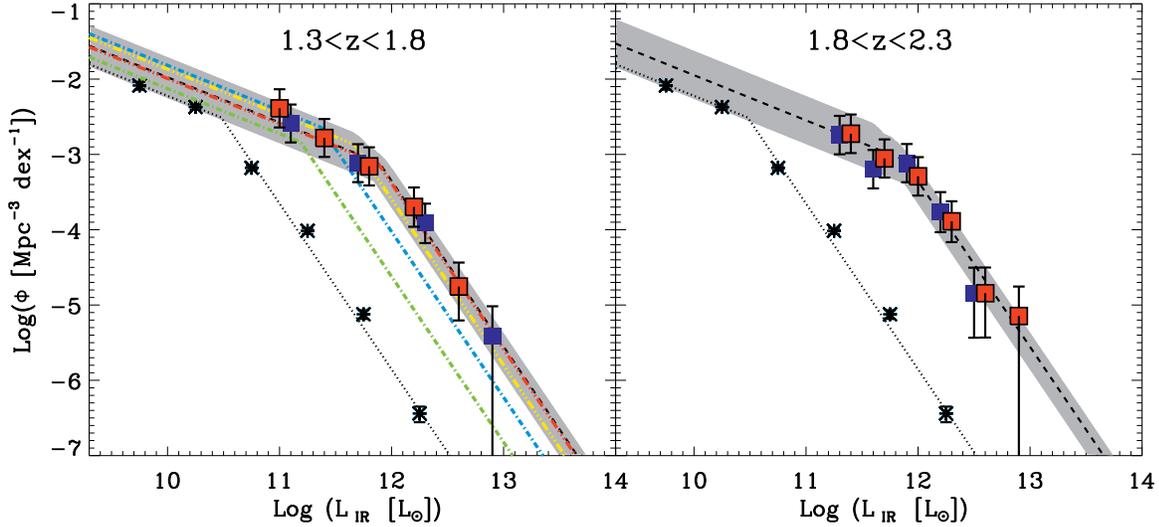}
	\caption{\label{fig:LF}Infrared LF estimated in two redshift bins with the $1/V_{max}$ method.
	Red and dark blue squares are obtained using $L_{IR}^{fit}$ and $L_{IR}^{70}$ respectively (see text).
	Asterisks show the local reference taken from \citet{sanders_2003} and the dotted line represents the best-fit to these data points with a double power law function with fixed slopes (see text).
	The dashed line represents the best fit of the infrared LF assuming that the shape of the infrared LF remains the same since $z\thicksim0$.
	The dark shaded area span all the solutions obtained with the $\chi^{2}$ minimization method and compatible, within 1 $\sigma$, with our data points.
	In the first redshift panel, we reproduce in green, blue, yellow and red the best fit of the LF obtained at $0.4<z<0.7$, $0.7<z<1.0$, $1.0<z<1.3$ \citep{magnelli_2009}, and $1.8<z<2.3$ respectively.
	}
\end{figure*}
\indent{
The reference catalog is constructed as follows.
We first start from a simulated catalog generated from the model of \citet{leborgne_2009} that best fits number counts of sources at 15, 24, 70, 160, 850 $\mu$m.
This catalog, which contains all infrared sources that should be observed over a field of 283 arcmin$^{2}$ with $0<z<5$, reproduces the observed infrared LF up to $z\thicksim1.3$ \citep[e.g., see figure 12 of][]{magnelli_2009}.
We only keep sources with $0<z<1.3$ or $z>2.3$.
Then from our observed infrared LF (section \ref{sec:discussion}) we construct and add to this catalog all infrared sources  that should be observed at $1.3<z<2.3$ over a field of 283 arcmin$^{2}$.
This catalog, which by construction reproduces all the observed infrared LF from $z=0$ up to $z\thicksim2.3$, will be our reference catalog.
\\}
\indent{
Starting from this reference catalog we create 1000 mock catalogs which contain the same number of sources as the original one but we attribute to each source a new redshift randomly selected to reproduce the observed distribution of spectroscopic versus photometric redshifts (see Figure \ref{fig:data}).
To take into account the bolometric correction uncertainties not associated with the photometric redshift errors, we attribute to each source a new infrared luminosity selected inside a Gaussian distribution centered at the original source luminosity and with a dispersion of 0.2 dex (see Section \ref{subsec:LIR}).
Using these 1000 mock catalogs, we then compute the infrared LF and study the difference between these infrared LFs and the infrared LF derived from the reference catalog.
\\ \\}
\indent{
Using these Monte Carlo simulations, we find only small systematic offsets between the real infrared LF and the one inferred in presence of redshift uncertainties (see figure \ref{fig:monte}).
At $z\thicksim2$ and at faint luminosities, we underestimate the LF values by at most 0.1-0.15 dex while at bright luminosities we overestimate the LF values by at most 0.1-0.2 dex.
These systematic offsets are smaller than the total uncertainty in each luminosity bin ($\thicksim0.25$ dex) defined as the quadratic sum of the Poissonian error ($\propto\,1/\sqrt{N}$) and the dispersion given by the Monte Carlo simulations.
As a result, in the following we do not correct the inferred LF for these systematic offsets.
}
\subsection{Results\label{sec:discussion}}
\indent{
In figure \ref{fig:LF}, we present the infrared LF derived in two redshift bins ($1.3<z<1.8$ and $1.8<z<2.3$) using our two different infrared bolometric corrections (\textit{red} and \textit{blue squares} for $L_{IR}^{fit}$ and $L_{IR}^{70}$ respectively, see Table \ref{tab:LF IR} and  \ref{tab:LF IR bis}).
First, we note that the infrared LFs derived using these two different bolometric corrections are in very good agreement and certainly within the error bars.
Here on, we will refer to the LF derived using $L_{IR}^{fit}$ as the infrared LF.
Indeed we have a better understanding of the uncertainties of this technique and this bolometric correction was found to be reliable up to $z\thicksim2$ by \citet{murphy_2009}.
\\}
\indent{
We take as a local reference the infrared LF derived by \citet{sanders_2003}, showing their data points (\textit{stars}) and their best double power law fit (i.e., $\phi \propto L^{-0.6}$ for Log($L/L_{\odot}$)$<10.5$ and  $\phi \propto L^{-2.15}$ for Log($L/L_{\odot}$)$>10.5$).
Then, using a $\chi^{2}$ minimization, we fit our infrared LFs with the same function, fixing the power law slopes at their $z\thicksim0$ values and leaving $L_{knee}$ and $\phi_{knee}$ as free parameters (see Table \ref{tab:fit parameter}).
The shaded regions present all the solutions which are compatible with the data within 1 $\sigma$.
These shaded regions extend to luminosities lower than our current observations and strongly depend on the assumption we made on the shape of the infrared LF, i.e. that it remains the same since $z\thicksim0$.
We will see latter on that this assumption is quite consistent with observational constraints obtained on the low luminosity end of the infrared LF \citep{reddy_2008}.
\\}
\begin{figure}
	\includegraphics[width=9cm]{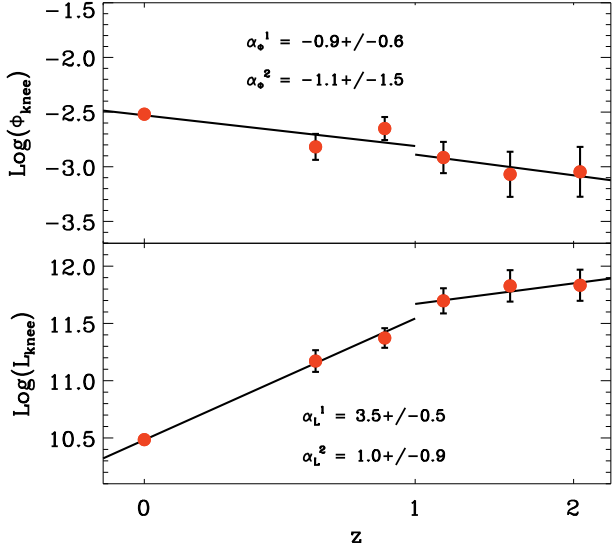}
	\caption{\label{fig:LF evol}
	Evolution of $\phi_{knee}$ and $L_{knee}$ as function of redshift.
	Data points below $z=1.3$ are taken from \citet{magnelli_2009}.
	}
\end{figure}
\begin{figure}
	\includegraphics[width=9.cm]{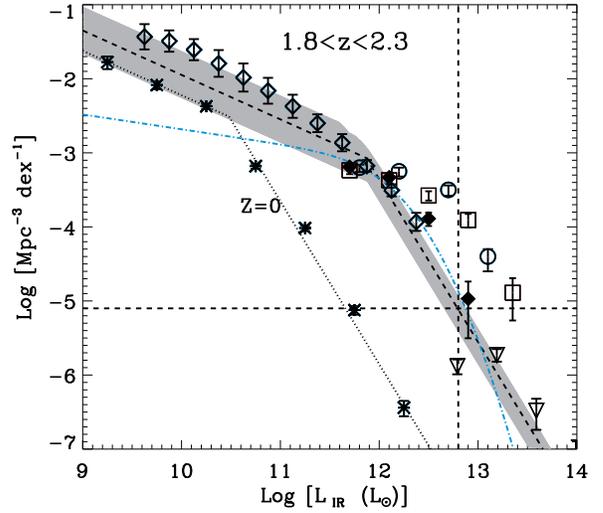}
	\caption{\label{fig:LF compa}
	The infrared LF at $z\thicksim2$ obtained in this work (\textit{dark shaded area} and \textit{dashed line}) as compared with the determinations of other authors.
	The infrared LF at $z\thicksim2$ obtained by \citet[][we are using their double exponential function]{caputi_2007} is represented by the dashed dotted line.
	Empty circles represent the infrared LF at $z\thicksim2$ inferred by \citet{perez_2005}.
	Empty diamonds represent the infrared LF at $z\thicksim2.3$ inferred by \citet{reddy_2008}.
	Empty triangles represent the infrared LF at $z\thicksim2.5$ inferred by \citet{chapman_2005}.
	Filled black diamonds and empty squares represent the infrared LF that we would have inferred using our 24 $\mu$m sample and the unscaled LDP or CE01 SED libraries respectively.
	The horizontal dashed line presents the source density below which the number of sources in the volume of GOODS and in a luminosity bin of 0.5 dex is less than 2.
	The vertical dashed line represents the corresponding luminosity using the best fit of our infrared LF. 
	}
\end{figure}
\begin{table*}
\caption{\label{tab:fit parameter}Parameter values of the infrared LF}
\centering
\begin{tabular}{ccccc}
\hline \hline
{Redshift} &{$\alpha_{1}\,^\mathrm{a}$} &{$\alpha_{2}\,^\mathrm{a}$} &{\rm{Log($L_{knee}$)}} & {\rm{Log($\phi_{knee}$)}} \\
& & &  {\tiny{\rm{Log($L_{\odot}$)}}} & {\tiny{\rm{Log($Mpc^{-3}dex^{-1}$)}}} \\
\hline
$z\thicksim0$ & $-0.60$ & $-2.20$ & $10.48\pm0.02 $ & $-2.52\pm0.03$\\
$0.4<z<0.7\,^\mathrm{b}$ & $-0.60$ & $-2.20$ & $11.19\pm0.04$ & $-2.84\pm0.06$\\
$0.7<z<1.0\,^\mathrm{b}$ & $-0.60$ & $-2.20$ & $11.37\pm0.03$ & $-2.65\pm0.05$\\
$1.0<z<1.3\,^\mathrm{b}$ & $-0.60$ & $-2.20$ & $11.69\pm0.06 $ & $-2.91\pm0.10$\\
$1.3<z<1.8$ & $-0.60$ & $-2.20$ & $11.84\pm0.13$ & $-3.07\pm0.20$\\
$1.8<z<2.3$ & $-0.60$ & $-2.20$ & $11.83\pm0.13 $ & $-3.04\pm0.22$\\
\hline
\end{tabular}
\begin{list}{}{}
\item[$^{\mathrm{a}}$] Fixed slopes
\item[$^{\mathrm{b}}$] These parameter values are taken from \citet{magnelli_2009}
\end{list}
\end{table*}
\indent{
The evolution of $L_{knee}$ and $\phi_{knee}$ between $z=1.3$ and $z=2.3$ is compared in Figure \ref{fig:LF evol} with the evolution found at lower redshifts by \cite{magnelli_2009}.
Assuming that the shape of the LF remains the same since $z\thicksim0$, we express the evolution of the infrared LF as $\rho(L,z)=g(z)\rho(L/f(z),0)$, where $g(z)$ and $f(z)$ describe the density and the luminosity evolution through $g(z)=(1+z)^{p}$ and $f(z)=(1+z)^{q}$.
Between $z=0$ and $z\thicksim1$ the redshift evolution consists mainly of a slight density evolution proportional to $(1+z)^{-0.9\pm0.6}$ and a luminosity evolution proportional to $(1+z)^{3.5\pm0.5}$.
Then between $z\thicksim1$ and $z\thicksim2$ we observe a density evolution proportional to $(1+z)^{-1.1\pm1.5}$ associated with a luminosity evolution proportional to $(1+z)^{1.0\pm0.9}$.
In comparison, at $z > 1$ \citet{caputi_2007} find a density evolution proportional to $(1+z)^{-3.9\pm1.0}$ and a luminosity evolution proportional to $(1+z)^{2.2\pm0.5}$.
The evolution of the infrared LF that we find at $z > 1.3$ is more gradual than that derived by \citet{caputi_2007} and is nearly consistent with no evolution.
\\ \\}
\indent{
Appendix \ref{sec:apA} presents the evolution of the rest-frame 8, 15, 25, 35 $\mu$m LFs.
Rest-frame luminosities of each source are derived using the SED that we used to compute its $L_{IR}^{fit}$.
\\}
\subsection{Comparison with previous work}
\indent{
In Figure \ref{fig:LF compa} we compare our results at $z\thicksim2$ with the infrared LF inferred in various previous studies.
There is a clear disagreement between our results and the infrared LF derived by \citet{caputi_2007}.
This discrepancy arises from the fact that the bolometric corrections used in \citet[][i.e. with the LDP library]{caputi_2007} do not take into account the SED evolution that we observe at high redshift.
Using the same bolometric corrections as \citet{caputi_2007} does lead us to similar results (see \textit{black diamonds} of Figure \ref{fig:LF compa}).
The disagreement of our results with the LF from \citet{perez_2005} is even larger because they compute their bolometric corrections using the CE01 library (see \textit{open squares} in Figure \ref{fig:LF compa}), i.e. the standard SED library which exhibit the largest discrepancies with the observed $L_{24\,\mu \rm{m/(1+z)}}-L_{70\,\mu \rm{m/(1+z)}}$ correlation at $z > 1.3$ (see Section \ref{subsec:LIR}).
\citet{rodighiero_2010} have also derived the $z\thicksim2$ infrared LF using deep 24~$\mu$m observations of the VVDS-SWIRE ($S_{24\ \mu\rm m}>400\ \mu$Jy) and GOODS ($S_{24\ \mu\rm m}>80\ \mu$Jy) fields.
Data points from this study are not shown in Figure \ref{fig:LF compa} since they are very similar to that from \citet{caputi_2007} \citep[see figure 15 of][]{rodighiero_2010}.
\\}
\indent{
In Figure \ref{fig:LF compa}, we also compare our results with the infrared LF inferred at $z\thicksim2.3$ by \citet{reddy_2008} using observations of UV-selected star-forming galaxies.
This study derived SFR and dust reddening from the UV rest-frame observations calibrated by comparison to 24~$\mu$m photometry for brighter sources.
The UV-derived extinction was used to compute the expected infrared emission from galaxies fainter than the 24~$\mu$m detection limit and hence provide an extension of the IR LF to fainter luminosities.
We find good agreement between our LF and that of Reddy et al.\ in the range of luminosities where the two studies overlap.
At faint luminosities ($\log (L_{IR}/L_\odot) < 11$) our best fit LF falls somewhat below that derived by \citet{reddy_2008}, although they are still consistent within the uncertainties.
We have no direct measurements with \textit{Spitzer} at such faint luminosities and rely upon an extrapolation based on a faint-end slope fixed at its $z \sim 0$ value.
As detail in section \ref{sec:SFH} this disagreement at faint luminosities has nearly no impact on the SFR density inferred at $z\thicksim2$.
Indeed, the integrated SFR density of the universe at $z\thicksim2$ computed from our LF agrees with that derived by \citet{reddy_2008}.
\\}
\indent{
Finally we compare our results with the infrared LF inferred at $z\thicksim2.5$ by \citet{chapman_2005} using submillimeter observations\footnote{S.\ Chapman (private communication) confirms that the far-infrared luminosities for the submm galaxy LF reported in Table 6 of \citet{chapman_2005} are integrated over the wavelength range 8-1100~$\mu$m, nearly the same as the range 8-1000~$\mu$m that we adopt here.
He also verifies that the luminosity column of that table should include an unspecified factor of $h^2$.
We have converted the data from Chapman et al.\ (2005) Table 6 to the value $H_0 = 70$ km s$^{-1}$ Mpc$^{-3}$ that we adopt in this paper.}.
The luminosity range probed by \citet{chapman_2005} is not constrained by our study since the comoving volume probed by the GOODS survey is too small (i.e., fewer than 2 sources would be present in this volume for a luminosity bin of $\Delta Log(L_{IR})=0.5$ dex).
We note however that the extrapolation of our infrared LF to high luminosities is consistent with the estimates of \citet{chapman_2005}.
\\ \\}
\subsection{Discussion \label{sec:SFH}}
\begin{figure*}
\centering
	\includegraphics[width=8.2cm]{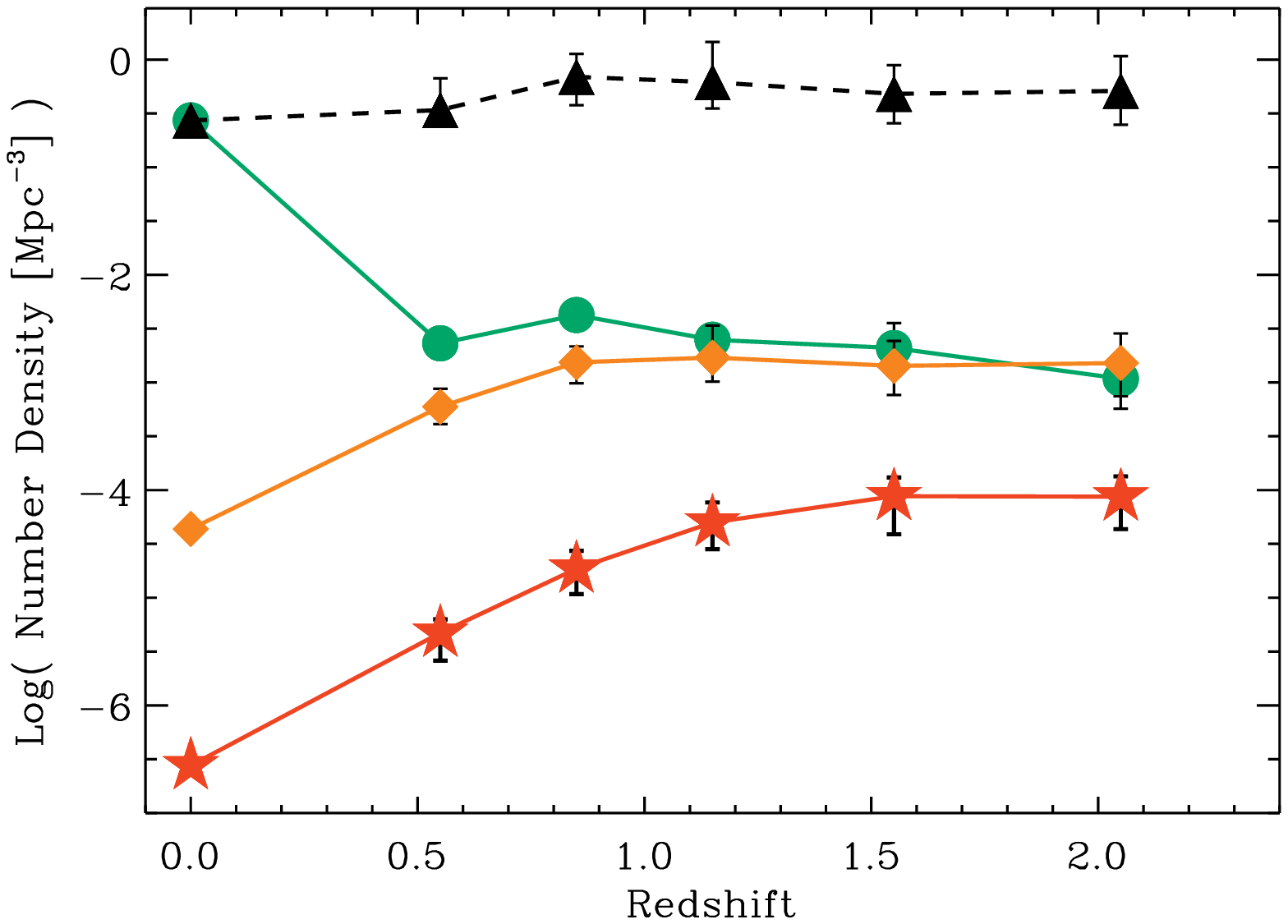}
	\includegraphics[width=8.5cm]{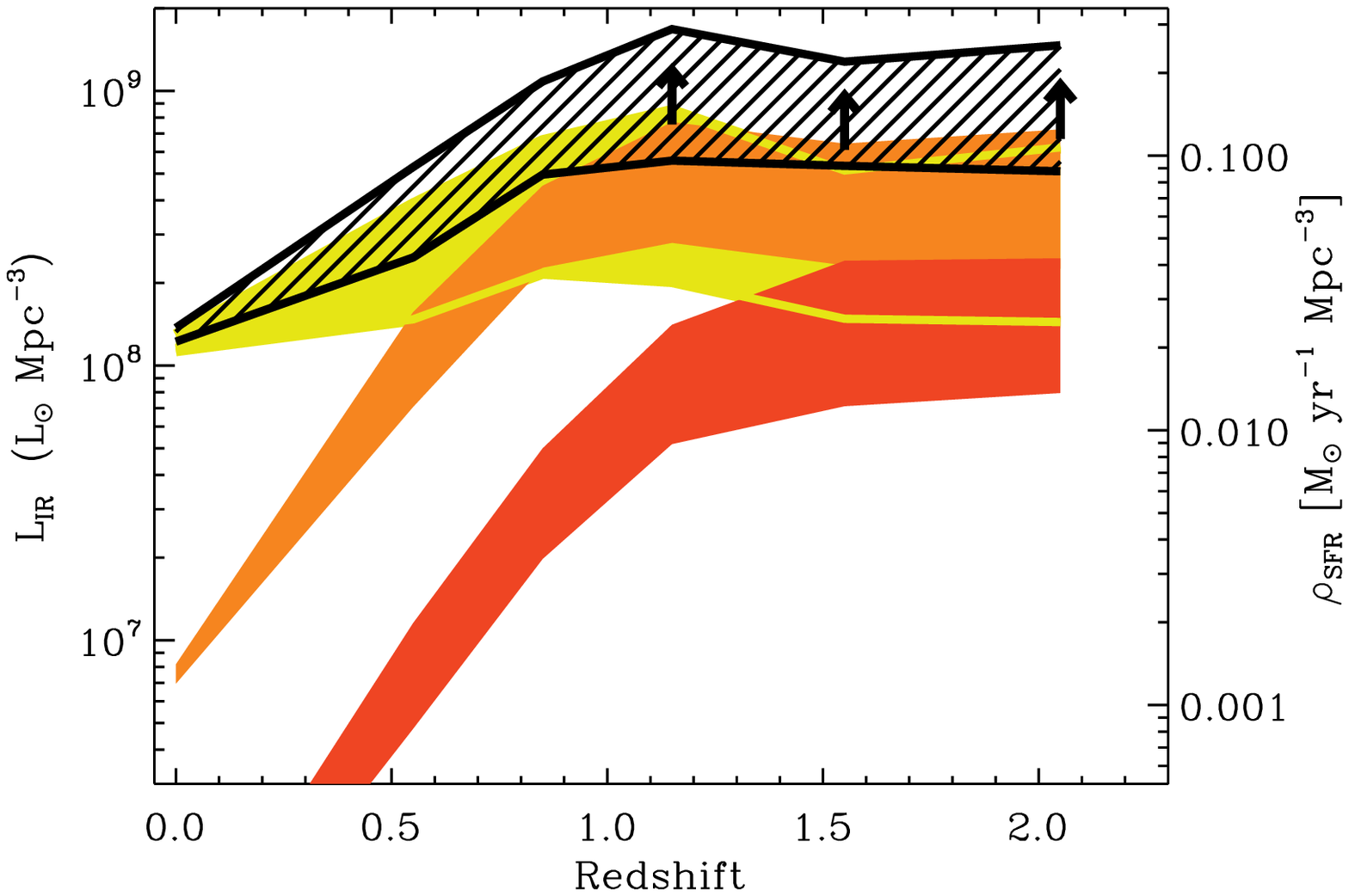}
	\caption{\label{fig:densite}\textbf{\textit{(Left)}} Evolution up to $z\thicksim2.3$ of the comoving number density of ``normal" galaxies (i.e. $10^{7}\rm{\,L_{\odot}}<\rm{L_{ir}}<10^{11}\,\rm{L_{\odot}}$; \textit{black filled triangles}), LIRGs (\textit{orange filled diamonds}) and ULIRGs (\textit{red filled stars}).
	The green circles represent the total number of galaxies which are above the 24~$\mu$m detection limit of the surveys presented here, i.e. $\rm{L_{ir}}>L_{ir}^{flux\ limit}$.
	The z$\,\thicksim0$ points are taken from \citet{sanders_2003}.
	\textbf{\textit{(Right)}} Evolution of the comoving IR energy density up to $z\thicksim2.3$ (\textit{upper striped area}) and the relative contribution of ``normal" galaxies (\textit{yellow filled area}), LIRGs (\textit{orange filled area}) and ULIRGs (\textit{red filled area}).
	The areas are defined using all the solutions compatible within 1 $\sigma$ with the infrared LF.
	Black arrows show the comoving IR energy density derived by stacking the 70 $\mu$m image at all IRAC sources positions ($S_{3.6\,\mu m}^{IRAC}>0.6\,\mu$Jy).
	The axis on the right side of the diagram shows the evolution of  the SFR density under the assumption that the SFR and L$_{\rm IR}$ are related by Eq.~\ref{eq:SFR_Lir} for a Salpeter IMF.
	}
\end{figure*}
\begin{figure*}
	\includegraphics[width=9.cm]{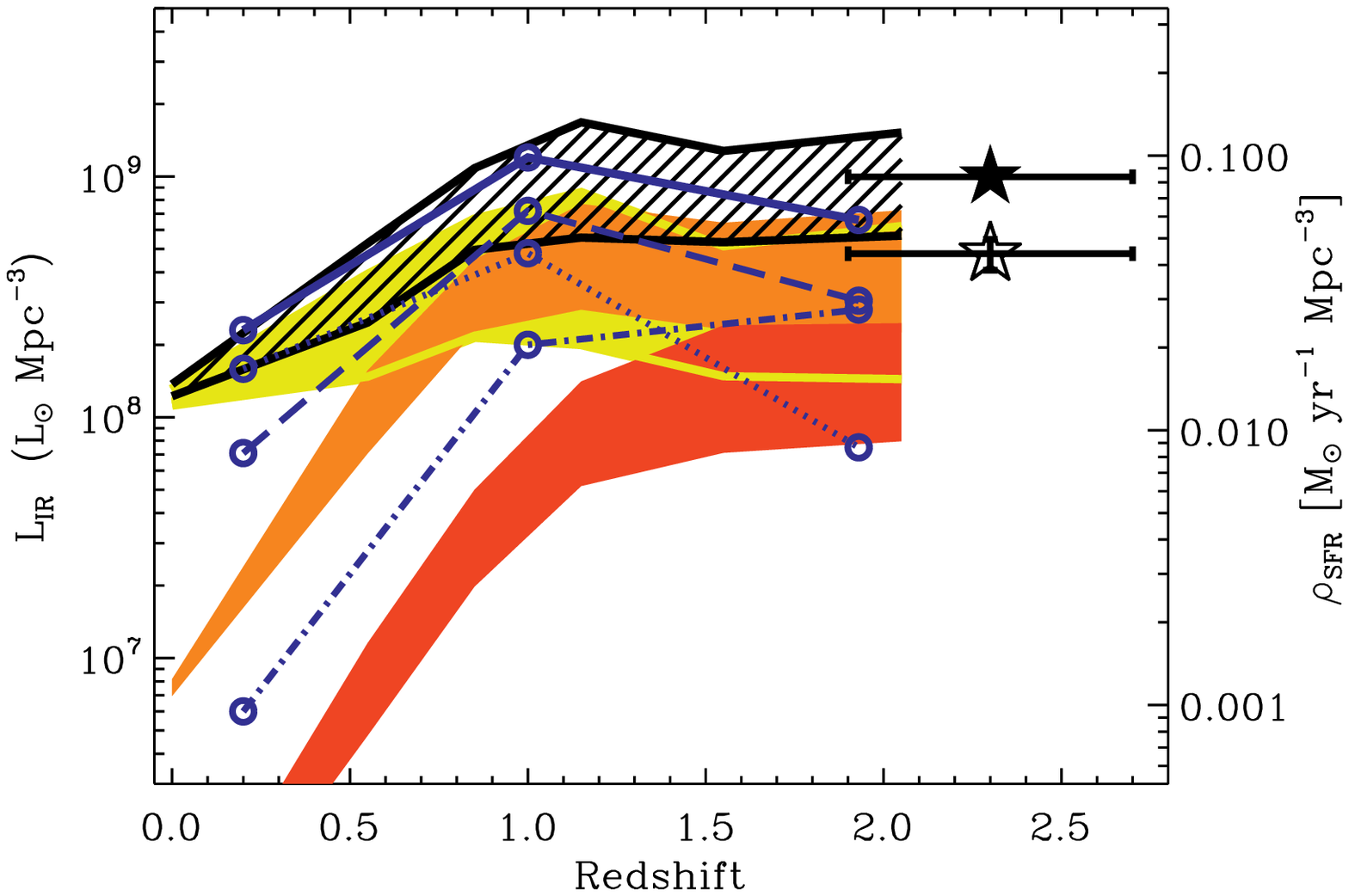}
		\includegraphics[width=9.cm]{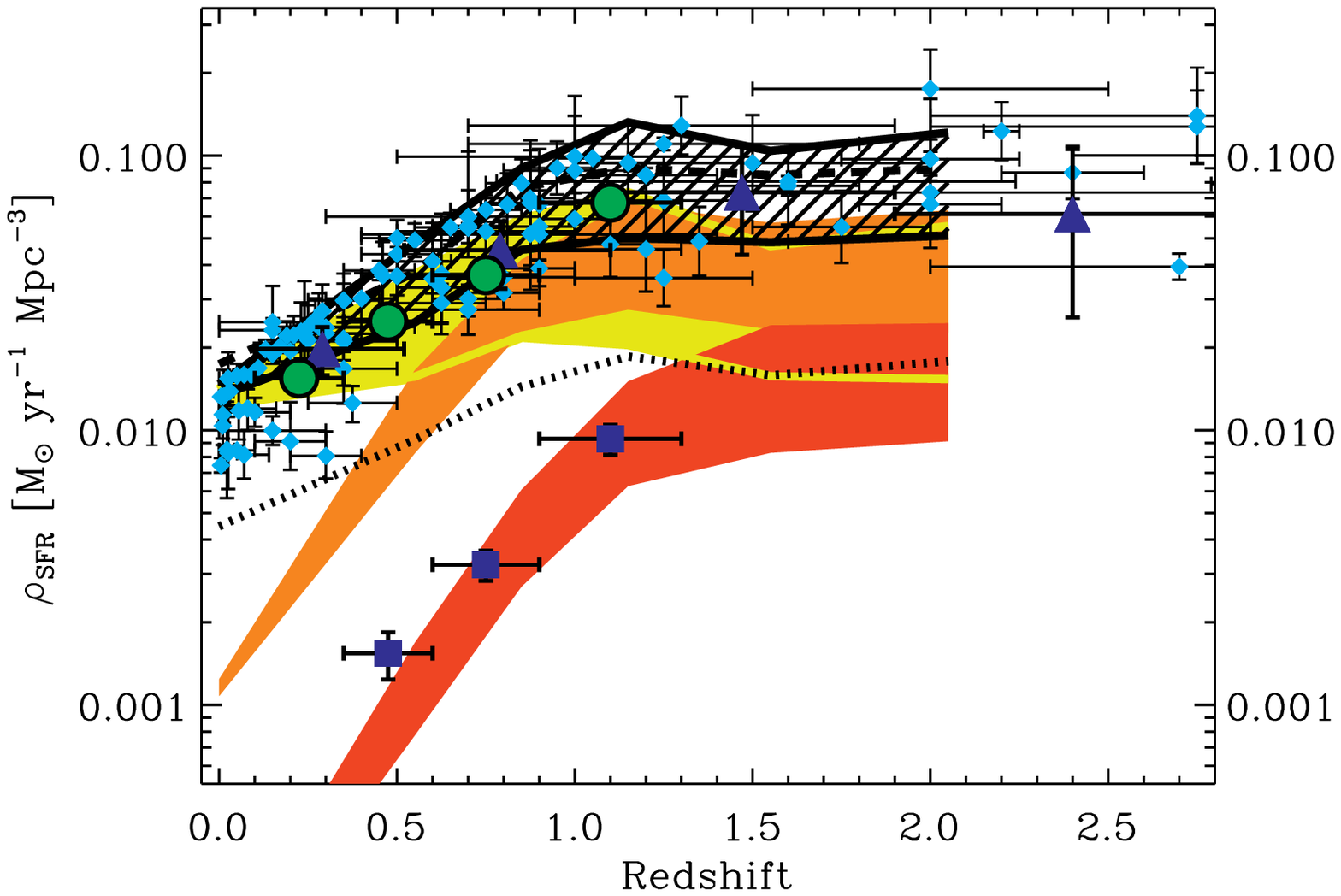}
	\caption{\label{fig:densite bis} \textbf{\textit{(Left)}} Evolution of the comoving IR energy density up to $z\thicksim2.3$. 
	Blue empty circles represent the results obtained by \citet{caputi_2007} for the global evolution of the comoving energy density (\textit{solid line}) and the relative contribution of ``normal" galaxies (\textit{dot line}), LIRGs (\textit{dashed line}) and ULIRGs (\textit{dot dashed line}).
	Filled black star represents the comoving IR energy density of the Universe inferred at $z\thicksim2.3$ by \citet{reddy_2008} while open star shows the relative contribution of LIRGs.
	Filled areas are as in Figure \ref{fig:densite}.
         \textbf{\textit{(Right)}} Evolution of the comoving SFR density up to $z\thicksim2.3$ assuming that SFR and L$_{\rm IR}$ are related by Eq.~\ref{eq:SFR_Lir} for a Salpeter IMF.
         	Filled areas are as in Figure \ref{fig:densite}.
	The dotted line represents the SFR measured using the UV light not corrected for dust extinction \citep{tresse_2007}.
	The dashed line represents the total SFR density defined as the sum of the SFR density estimated using our infrared observations and the SFR density obtained from the UV light uncorrected for dust extinction.
	Light blue diamonds are taken from \citet{hopkins_2006} and represent the SFR densities estimates using various estimators.
	Dark blue triangles represent the SFR density estimated by \citet{seymour_2008} using deep radio observations.
	Green circles represent the SFR density estimated by \citet{smolcic_2009} using deep 20 cm observations and dark blue squares represent the relative contribution of ULIRGs to this SFR density.
	}
\end{figure*}

\indent{
We derive the evolution of the comoving number density of LIRGs and ULIRGs by integrating the infrared LF at $z=1.55\pm0.25$ and $z=2.05\pm0.25$.
We then combine these estimates with the evolution found at $0<z<1.3$ by \citet{magnelli_2009} (Figure \ref{fig:densite} \textit{left}).
We find that the number densities of LIRGs and ULIRGs between $z\thicksim1.3$ and $z\thicksim2.3$ are nearly constant.
The number density of ULIRGs at $z \sim 2$ ($7.5^{+2.7}_{-4.0}\times10^{-5}\,$Mpc$^{-3}$) agrees with estimates made by \citet{daddi_2007a} ($\thicksim 10\times10^{-5}$ Mpc$^{-3}$) using UV observations calibrated against radio and other non-24~$\mu$m data.
However, if we compare our number density with the \citet{daddi_2007a} estimate obtained combining UV and 24~$\mu$m observations ($16^{+12}_{-6.0}\times10^{-5}\,$Mpc$^{-3}$) we find a clear disagreement.
\\}
\indent{
Figure \ref{fig:densite} (\textit{right}) presents the evolution of the comoving infrared luminosity density (IR LD; or, equivalently, SFR density under the assumption that the SFR and L$_{\rm IR}$ are related by Eq.~\ref{eq:SFR_Lir} for a Salpeter IMF, i.e. $\phi$(m)$\propto$ m$^{-2.35}$, between 0.1--100 M$_{\odot}$, Kennicutt \citeyear[][]{kennicutt_1998}) produced by ULIRGs, LIRGs, and by galaxies with $L_{IR} < 10^{11} L_\odot$ (hereafter called ``normal'' galaxies by analogy with ordinary spiral galaxies at $z\thicksim0$, although we note that at high redshift, LIRGs and ULIRGs themselves are sufficiently common to be considered ``normal'' for their epoch).
We find a slight decrease of the IR LD from $z=1.3$ to $z=2.3$ due to a decrease in the contribution of LIRGs and ``normal" galaxies.
At $z\thicksim2$ the IR LD of the Universe is still dominated by LIRGs and not by ULIRGs, contrary to previous claims \citep[e.g.,][]{perez_2005}.
Using the best fit of our infrared LF, we infer that at $z\thicksim2$ LIRGs and ULIRGs have an IR LD of $4.5\times10^{8}L_{\odot}{\rm\,Mpc^{-1}}$ and $1.5\times10^{8}L_{\odot}{\rm\,Mpc^{-1}}$ respectively and that they account for $49\%$ and $17\%$ of the total IR LD respectively.
\\}
\begin{equation}\label{eq:SFR_Lir} 
{\rm SFR}~ [{\rm M}_{\odot}~ {\rm yr}^{-1}] = 1.72 \times 10^{-10} L_{\rm IR} ~[{\rm L}_{\odot}]
\end{equation}
\indent{
We compare our estimates with the evolution derived by \citet{caputi_2007} (Figure \ref{fig:densite bis} \textit{left}).
At $z\thicksim2$ we find that the IR LD of ULIRGs estimated from our best fit is a factor of $\thicksim1.8$ lower than that estimated by \citet{caputi_2007}.
For LIRGs, for which we still have a good constraint due to our deep 24~$\mu$m sample, our estimate of their IR LD is a factor of $\thicksim1.5$ higher than that inferred by \citet{caputi_2007}.
Nevertheless, we note that if we take into account the range of IR LD defined by all solutions compatible within $1\sigma$ with our data points, our IR LDs of LIRGs and ULIRGs are compatible with estimates from \citet{caputi_2007}.
On the other hand, the IR LD estimated by \citet{caputi_2007} for galaxies with $L_{{\rm IR}}<10^{11}\ L_{\odot}$ is far below our estimate since they used a flatter faint-end slope for their infrared LF.
We believe that our estimate is more reliable since we are using a 24~$\mu$m catalog that is $\sim 3$ times deeper than that used by \citet{caputi_2007}.
We also note that the extrapolation of our infrared LF to these faint luminosities is corroborated by the infrared LF inferred by \citet{reddy_2008}.
\\}
\indent{
We also compare our $z\thicksim2.05$ IR LD values with the $z\thicksim2.3$ estimates of \citet{reddy_2008}.
We note that this comparison is not straightforward because their IR LD needs to be slightly corrected prior to be compared with our work.
Indeed, since they cannot constrain with their sample the contribution of ULIRGs to the IR LD, they use the value derived by \citet{caputi_2007}.
By replacing the \citet{caputi_2007} estimates by our value we compute the correct IR LD of \citet{reddy_2008}, i.e., $10.0\pm0.2\times10^{8}$ L$_{\odot}$ Mpc$^{-3}$.
We find that the IR LDs derived by \citet{reddy_2008} for LIRGs and for the integrated population of all galaxies agree with our values.
This agreement reinforces the idea that at $z\thicksim2$ UV corrected for extinction is a good SFR indicator \citep{daddi_2007a}.
\\ \\}
\indent{
In order to get a complete census on the SFR history we need to take into account the contribution of unobscured UV light.
The unobscured SFR density (\textit{dotted line} in Figure \ref{fig:densite bis} \textit{right}) is taken from \citet{tresse_2007} and corresponds to the SFR density inferred using UV observations not corrected for extinction.
The total SFR density (\textit{dashed line} in Figure \ref{fig:densite bis} \textit{right}) is then defined as the sum of the unobscured SFR density traced by the direct UV light and the dusty SFR density traced by the infrared emission.
We find that the relative contribution of unobscured UV light to the cosmic SFR density evolves nearly in parallel with the total one and accounts for $\sim$20\,\% of the total SFR density.
Globally, the cosmic star-formation history that we derived is consistent with the combination of indicators, either unobscured or corrected for dust extinction, as compiled by \citet{hopkins_2006}.
We also notice a very good agreement between the cosmic star-formation history derived in our work and the ones derived by \citet{seymour_2008} using deep radio observations.
\\ \\}
\indent{
Our estimates can suffer from several uncertainties.
Especially the contribution of ``normal" galaxies to the IR LD comes from the extrapolation to low luminosities of the infrared LF where we have no constraints.
To cross check our results we compute a lower limit on the comoving IR LD by stacking 70~$\mu$m images at the positions of all IRAC sources in each redshift bin of interest (i.e., $S_{3.6\,\mu m}^{IRAC}>0.6\,\mu$Jy; \textit{\textit{up arrows}} in Figure \ref{fig:densite} \textit{right}).
This analysis is possible because the correlation between $L_{70\mu{\mathrm m}/(1+z)}$ and $L_{IR}$ is quasi-linear at this redshift, and hence $\Sigma S(70\mu{\mathrm m}) \propto\ \Sigma L_{IR}$.
The stacking result is fully consistent with the value based on the integration of the extrapolated best fit to our infrared LF.
\\ \\}
\indent{
As discussed in Section \ref{subsec:LIR}, the role of obscured AGN on the estimate of the infrared LF is still uncertain.
Such results will be debated until the {\it Herschel} infrared space observatory provides accurate far-infrared measurements for faint, high-redshift galaxies.
However, as shown by \citet{murphy_2009} using IRS spectroscopy, the mid-IR continuum from an AGN appears to scale with increasing 24 $\mu$m luminosity.
As a result, the removal of any additional contribution from obscured AGN activity will only steepen the bright-end of the infrared LF.
This would reinforce our main result which is the fact that at $z\thicksim2$ previous studies have overestimated the number density of ULIRGs.
\\ \\}
\section{Conclusion \label{sec:conclusion}}
\indent{
For the first time we take advantage of deep far-infrared observations to derive the evolution of the infrared luminosity density (or equivalently SFR density under the assumption that the SFR and L$_{\rm IR}$ are related by Eq.~\ref{eq:SFR_Lir} for a Salpeter IMF) over the last 4/5ths of the cosmic time (see Figure \ref{fig:densite} \textit{right}).
Using the deepest 24~$\mu$m ($\thicksim3$ times deeper than any previous studies) and 70~$\mu$m observations made by \textit{Spitzer} and a careful stacking analysis we are able to calibrate a new infrared bolometric correction based on the renormalization of local SED templates reproducing the observed $L_{24/(1+z)}$ vs $L_{70/(1+z)}$ correlations.
This new bolometric correction is a key result of our analysis, since previous studies on the infrared LF at $z \sim 2$ \citep[e.g.,][]{perez_2005,caputi_2007} did not account for the evolution of the infrared SEDs that we observe in high-redshift galaxies.
Our main result is that we find a flattening of the SFR density between $z = 2.3$ and 1.3, and that the comoving number density of LIRGs and ULIRGs remain nearly constant over this redshift range.
At $z\thicksim2$ the SFR density is still dominated by LIRGs and not by ULIRGs contrary to previous claims.
The flattening of the SFR density at $z>1$ reinforces the idea that at this redshift we observe a change of the properties of star-forming galaxies as already suggested by the reversal of the star formation-density relation at $z\thicksim1$ \citep{elbaz_2007}.
\\}
\indent{
The evolution of the SFR density of the Universe provides a strong constraint on the main mechanism which triggers the SFR in galaxies.
At $z<1$ the decrease of the SFR density might be driven by a gradual gas exhaustion as suggested by the continuous decrease of the SFR vs M$_{\star}$ relation in this redshift range \citep{noeske_2007b,elbaz_2007}.
Between $z \sim 2$ and $z \sim 1$, the relatively constant SFR density still needs to be understood in the framework of large-scale structure formation, merging, and/or AGN activity.
\\}
\indent{
The main limitation of our study is the uncertainty on the influence of obscured AGN on the infrared bolometric correction to be applied to bright 24 $\mu$m sources. 
This influence will soon be assessed using new far-infrared observations from \textit{Herschel}.  
In particular, the GOODS-Herschel Open Time Key Programme (PI: David Elbaz) will reach the faintest flux limits at 100~$\mu$m in an ultradeep field within GOODS-S, expected to provide individual measurements for most $z \sim 2$ galaxies detected at 24~$\mu$m by \textit{Spitzer}, where here we could only derive average values based on 70~$\mu$m stacking.
This should help disentangle the contributions of AGN and star formation for sources over a broad swath of the high-redshift infrared luminosity function.
}
\acknowledgements{
We would like to thank Andrea Cimatti for permission to use the GMASS redshifts and Daniel Stern and Hyron Spinrad for permission to use their GOODS Keck redshifts.
This work is based on observations made with the Spitzer Space Telescope, which is operated by the Jet Propulsion Laboratory, California Institute of Technology under a contract with NASA. 
Support for this work was provided by NASA through an award issued by JPL/Caltech.
B. Magnelli would like to thank Scott Chapman for clarifying issues about the IR LF for submm galaxies from his 2005 paper.
D.Elbaz wishes to thank the Centre National d'Etudes Spatiales (CNES) for their support.
}
 \bibliographystyle{aa}

\Online
\begin{appendix}
\section{The Rest-frame 8 $\mu$m, 15 $\mu$m, 25 $\mu$m and 35 $\mu$m LFs\label{sec:apA}}
\begin{figure*}
\centering
	\includegraphics[width=18cm]{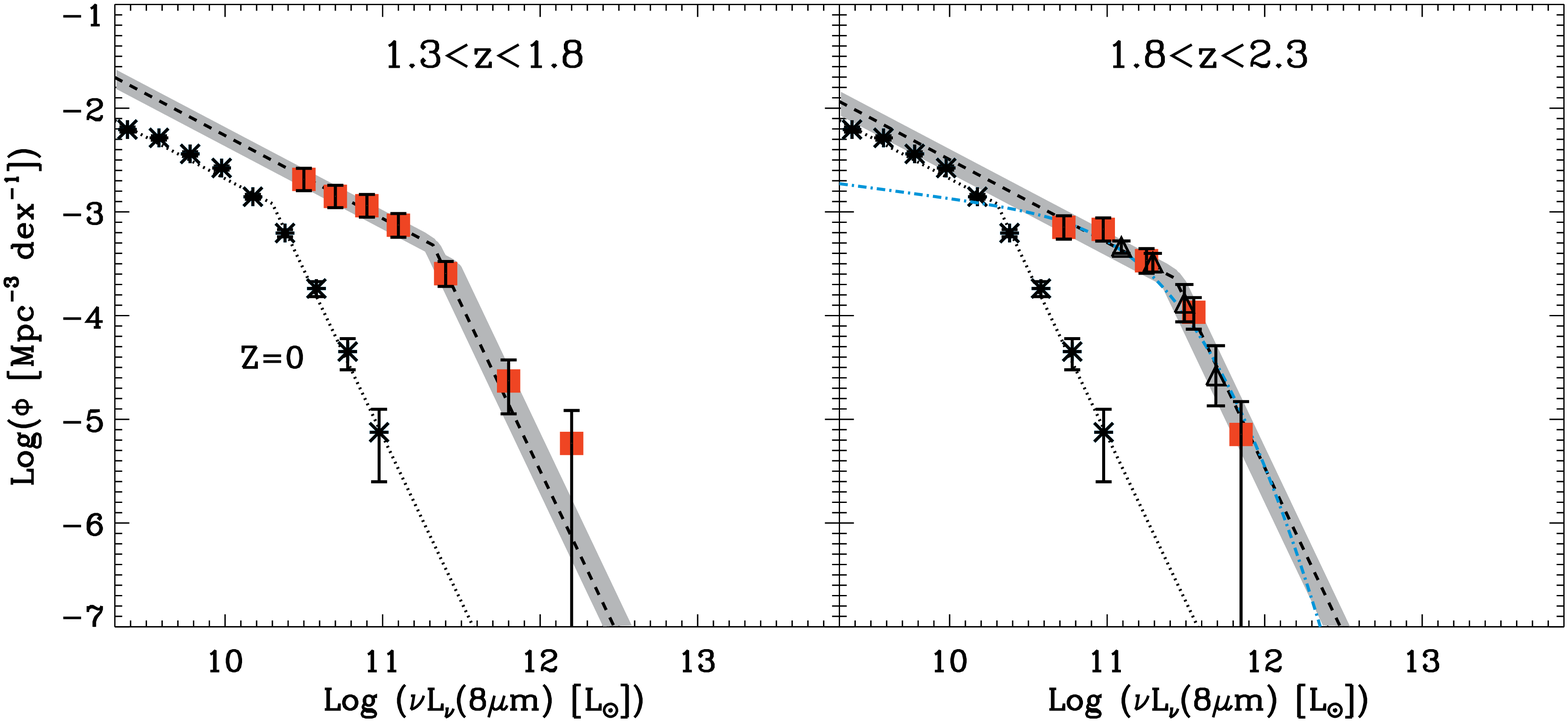}
	\caption{\label{fig:LF 8}The rest-frame 8 $\mu$m LF estimated in two redshift bins with the $1/V_{max}$ method.
	Red squares are obtained using scaled CE01 templates which best fit the $L_{24\,\mu \rm{m/(1+z)}}-L_{70\,\mu \rm{m/(1+z)}}$ correlation.
	Empty triangles and blue dashed-dotted line present the rest-frame 8 $\mu$m LF obtained at $z\thicksim2$ by \citet{caputi_2007}.
	Asterisks show the local reference taken from \citet{huang_2007} and the dotted line presents the best-fit to these data points with a double power law function with fixed slopes (see Table \ref{tab:fit parameter tot}).
	The dark shaded area span all the solutions obtained with the $\chi^{2}$ minimization method and compatible, within 1 $\sigma$, with our data points.
	The dashed line represents the best fit of the rest-frame 8 $\mu$m LF.
	}
\end{figure*}

\begin{figure*}
\centering
	\includegraphics[width=18cm]{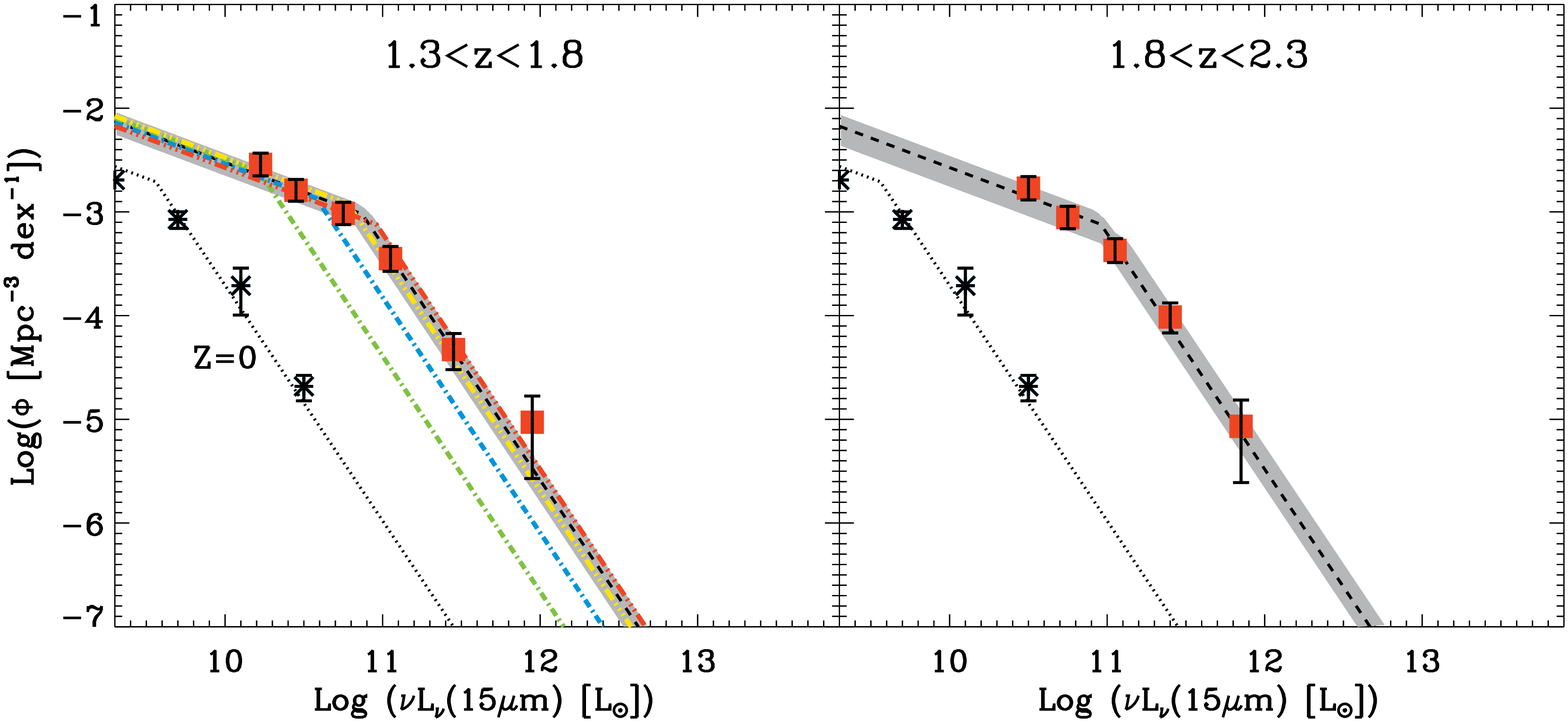}
	\caption{\label{fig:LF 15}The rest-frame 15 $\mu$m LF estimated in two redshift bins with the $1/V_{max}$ method.
	Red squares are obtained using scaled CE01 templates which best fit the $L_{24\,\mu \rm{m/(1+z)}}-L_{70\,\mu \rm{m/(1+z)}}$ correlation.
	Asterisks show the local reference taken from \citet{xu_2000} and the dotted line presents the best-fit to these data points with a double power law function with fixed slopes (see Table \ref{tab:fit parameter tot}).
	The dark shaded area span all the solutions obtained with the $\chi^{2}$ minimization method and compatible, within 1 $\sigma$, with our data points.
	The dashed line represents the best fit of the rest-frame 15 $\mu$m LF.
	In the first redshift panel, we reproduce in green, blue, yellow and red the best fit of the LF obtained at $0.4<z<0.7$, $0.7<z<1.0$, $1.0<z<1.3$ \citep{magnelli_2009}, and $1.8<z<2.3$ respectively.
	}
\end{figure*}

\begin{figure*}
\centering
	\includegraphics[width=18cm]{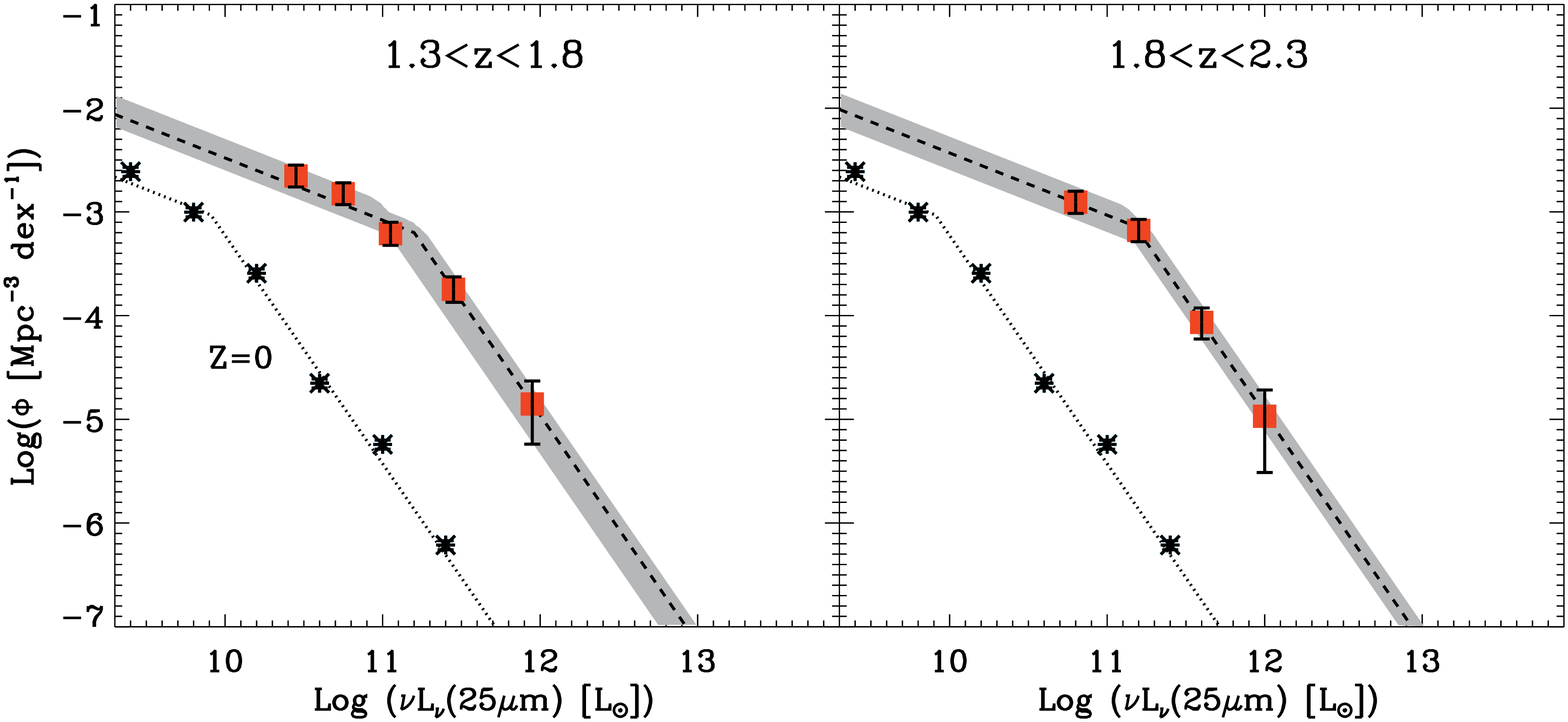}
	\caption{\label{fig:LF 25}The rest-frame 25 $\mu$m LF estimated in two redshift bins with the $1/V_{max}$ method.
	Red squares are obtained using scaled CE01 templates which best fit the $L_{24\,\mu \rm{m/(1+z)}}-L_{70\,\mu \rm{m/(1+z)}}$ correlation.
	Asterisks show the local reference taken from \citet{shupe_1998} and the dotted line presents the best-fit to these data points with a double power law function with fixed slopes (see Table \ref{tab:fit parameter tot}).
	The dark shaded area span all the solutions obtained with the $\chi^{2}$ minimization method and compatible, within 1 $\sigma$, with our data points.
	The dashed line represents the best fit of the rest-frame 25 $\mu$m LF.
	}
\end{figure*}

\begin{figure*}
\centering
	\includegraphics[width=18cm]{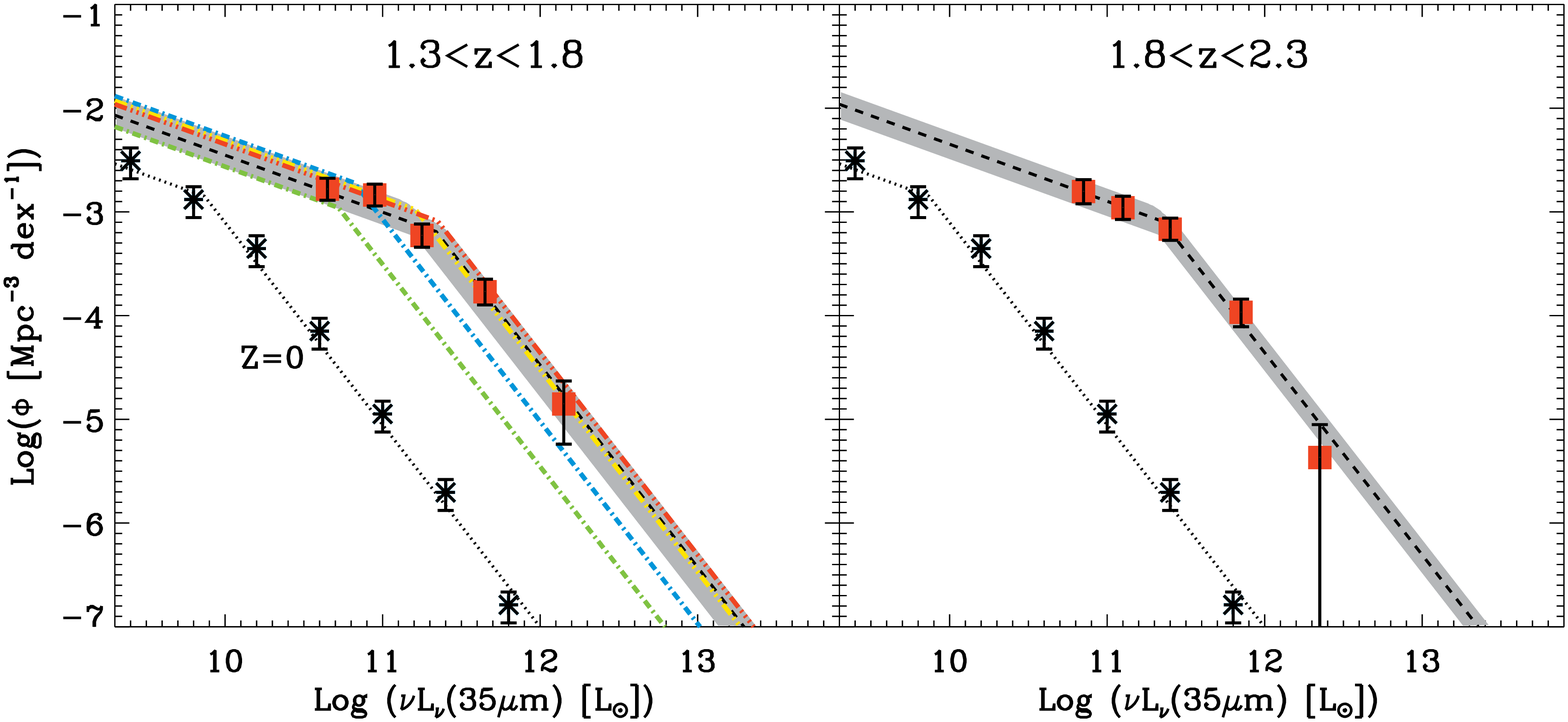}
	\caption{\label{fig:LF 35}The rest-frame 35 $\mu$m LF estimated in two redshift bins with the $1/V_{max}$ method.
	Red squares are obtained using scaled CE01 templates which best fit the $L_{24\,\mu \rm{m/(1+z)}}-L_{70\,\mu \rm{m/(1+z)}}$ correlation.
	Asterisks show the local reference derived from \citet{shupe_1998} and the dotted line presents the best-fit to these data points with a double power law function with fixed slopes (see Table \ref{tab:fit parameter tot}).
	The dark shaded area span all the solutions obtained with the $\chi^{2}$ minimization method and compatible, within 1 $\sigma$, with our data points.
	The dashed line represents the best fit of the rest-frame 35 $\mu$m LF.
	In the first redshift panel, we reproduce in green, blue, yellow and red the best fit of the LF obtained at $0.4<z<0.7$, $0.7<z<1.0$, $1.0<z<1.3$ \citep{magnelli_2009}, and $1.8<z<2.3$ respectively.
	}
\end{figure*}

\indent{
We aim to derive the rest-frame 8 $\mu$m, 15 $\mu$m, 25 $\mu$m and 35 $\mu$m LF  from our 24 $\mu$m sample.
This is done here for several reasons.
While the derivation of the rest-frame 8 $\mu$m LF has already been addressed in some previous studies \citep{caputi_2007}, our 24~$\mu$m sample reaches flux limits $\sim 3$ times fainter, providing improved constraints on the LF break.
The rest-frame 15~$\mu$m LF provides continuity with what we have computed in \citet{magnelli_2009}.
The rest-frame 25 $\mu$m LF was not derived in \citet{magnelli_2009} but it does have several points of interest.
First, it reduces the $k$--correction that one has to apply when observing infrared sources at $z\thicksim2$ using a 70 $\mu$m passband or at $z\thicksim3$ using a 100 $\mu$m passband.
Hence, this rest-frame LF could be compared with future 70 $\mu$m and 100 $\mu$m observations made using the Photodetector Array Camera and Spectrometer (PACS) instrument onboard the \textit{Herschel} satellite.
Second, the rest-frame 25~$\mu$m LF offers a means to compare directly with IRAS 25~$\mu$m observations of galaxies in the local universe.
Third, and perhaps most importantly, the rest-frame 25~$\mu$m luminosity has been shown to provide a reliable and fairly direct measurement of star formation in galaxies \citep[e.g.,][]{calzetti_2007}.
Hence, one may want to use these measurements in the future to directly derive the SFR distribution function without using intermediate bolometric corrections.
The rest frame 35~$\mu$m LF also provides continuity with quantities derived in \citet{magnelli_2009} and corresponds at $z\thicksim2$ to the observed 100~$\mu$m flux density.
As a result, this rest-frame LF would be a standard comparison for the $z\thicksim2$ LF computed using PACS 100 $\mu$m observations.
\\}
\indent{
The various rest-frame luminosities are derived using the same method as that used to compute $L_{IR}^{fit}$.
For each 24 $\mu$m source we deduce its 70 $\mu$m flux density using the $L_{24\,\mu \rm{m/(1+z)}}-L_{70\,\mu \rm{m/(1+z)}}$ correlation.
Then we choose in the CE01 library the scaled template which best fits these 24 and 70 $\mu$m fluxes densities.
Using this scaled template we then compute the rest-frame luminosities of this galaxy in our four passbands of interest (i.e., at 8 $\mu$m, 15 $\mu$m, 25 $\mu$m and 35~$\mu$m).
The corresponding rest-frame LFs are computed using the $1/V_{max}$ method.
All the LF are then fitted using a double power law function with fixed slopes estimated using a bivariate method (see Table \ref{tab:fit parameter tot}).
\\}
\indent{
Figure \ref{fig:LF 8} presents the rest-frame 8 $\mu$m LF derived in our two redshift bins using the $1/V_{max}$ method.
These LF are compared with the local reference of \citet{huang_2007} and with the $z\thicksim2$ LF derived by \citet{caputi_2007}.
Our estimates agree well with those of \citet{caputi_2007}, but extend to lower luminosities.
\\}
\indent{
Figure \ref{fig:LF 15} presents the rest-frame 15 $\mu$m LF derived in our two redshift bins using the $1/V_{max}$ method.
Comparing to the local reference of \citet{xu_2000} and with the $z\thicksim0.55$, $z\thicksim0.85$ and $z\thicksim1.15$ reference of \citet{magnelli_2009}, we note the strong evolution of this LF with redshift.
\\}
\indent{
Figure \ref{fig:LF 25} presents the rest-frame 25 $\mu$m LF derived in our two redshift bins using the $1/V_{max}$ method as well as the local reference of \citet{shupe_1998}.
\\}
\indent{
Finally Figure \ref{fig:LF 35} presents the rest-frame 35 $\mu$m LF derived in our two redshift bins using the $1/V_{max}$ method as well as the local reference derived from \citet{shupe_1998}, and the LFs derived in \citet{magnelli_2009} at $z\thicksim0.55$, $z\thicksim0.85$ and $z\thicksim1.15$.
\\}

\section{Source list}
\indent{
At the resolution of \textit{Spitzer} most of the sources in our fields are point sources (i.e FWHM$\thicksim5.9\arcsec$ and 18\arcsec at 24 $\mu$m and 70 $\mu$m respectively).
Therefore, to derive their photometry we decided to use a PSF fitting technique that take into account, as prior information, the expected position of the sources.
Starting from IRAC positions \citep[GOODS-N: GOODS legacy program, Dickinson et al., in preparation; GOODS-S: SIMPLE  catalog,][]{damen_2011} we extract all 24 $\mu$m sources.
Then, using our 24 $\mu$m catalogs, we extract all 70 $\mu$m sources.
This method deals with a large part of the blending issues encountered in dense fields and allows straightforward multi-wavelength association between near-, mid- and far-infrared sources.
The disadvantage of this method is that we have to assume that all sources present in our mid-infrared images have already been detected at IRAC wavelengths.
In our case this assumption is true because our IRAC 3.6~$\mu$m data are 30 times deeper than our current 24~$\mu$m observations and that the typical $S_{24\,\mu m}/S_{3.6\,\mu\rm{m}}$ ratio spans the range [2-20].
\\ \\}
\indent{
In this online material we release our complete 24~$\mu$m and 70~$\mu$m source catalogs for both GOODS fields.
These catalogs expend below the 80$\%$ completeness limit, and cover the full area (approximately $10\arcmin\times16\arcmin$) of the GOODS-S region (i.e., not only the smaller $10\arcmin\times10\arcmin$  region with the deepest 70 $\mu$m imaging that is used for the analysis in this paper).
The noise level in the GOODS 24 $\mu$m data is homogeneous over most of the field, with some degradation near the edge where the exposure time is somewhat reduced.
We restrict our release to regions with exposure time higher than $9500$ s per pixel.
This limit corresponds to a quarter of the exposure time of the deep inner region ($\thicksim38000$ s).
This degradation does not really affect the depth of our 24~$\mu$m catalogs in that region since uncertainties are still dominated by confusion ($\sigma^{map}_{edge}\thickapprox2\times\sigma^{map}_{inner}\thickapprox6\,{\rm \mu Jy}<\sigma^{confusion}\thickapprox7\,\mu$Jy).
At 70 $\mu$m, the noise level is roughly uniform throughout GOODS-N ($\thicksim12\,000$ s per pixel).
However, in GOODS-S the deepest 70 $\mu$m data, with noise similar to those in GOODS-N, are limited to a region approximately $10\arcmin\times10\arcmin$ in extent.
The outer region portions of the GOODS-S field have somewhat shallower 70 $\mu$m data ($\thicksim6\,000$ s per pixel).
\\ \\}
\indent{
At 24 $\mu$m, sources are detected using an empirical 24 $\mu$m PSF constructed with isolated point like objects present in the mosaic.
At 70 $\mu$m no reliable empirical PSF could be constructed because only a few isolated sources could be found in each map.
We then decided to use the appropriate 70 $\mu$m Point Response Function (PRF) estimated on the extragalactic First Look Survey mosaic \citep[xFLS;][]{frayer_2006_b} and available on the \textit{Spitzer} web site.
At both wavelengths an aperture correction is applied to all flux densities to account for the finite size of our PSFs.
Those aperture corrections are taken from the \textit{Spitzer} data handbook.
\\}
\indent{
Calibration factors used to generate the final 24 and 70 $\mu$m mosaics are derived from stars, whose SED at these wavelengths are generally very different from those of distant galaxies.  
Hence, color-corrections have to be applied to all flux densities (at most $\thicksim 10\%$).
In the catalogs released here, 70~$\mu$m flux densities have been color-corrected using a systematic and standard correction of $1.09$ (see \textit{Spitzer} data handbook).
This 70 $\mu$m color-correction is computed for distant galaxies with dust temperature of $\thicksim40$ K.
This color-correction differs from those applied in our study and which take into account the redshift of each source (see discussion in Section \ref{subsec:infrared data}).
No color-corrections are applied to our 24 $\mu$m flux densities since, for those data, color-corrections are more strongly dependent on the redshift of the source.
Indeed, 24 $\mu$m data probes different part of galaxy SED as function of the redshift (black-body emission of dust or PAH emission).
\\ \\}
\indent{
Our 24 $\mu$m and 70 $\mu$m data are the deepest observations taken by \textit{Spitzer} and have been designed to reach the confusion limit of this satellite.
Flux uncertainties are therefore a complex combination of photon and confusion noise.
In order to estimate these complex flux uncertainties and to characterize the quality of our 24 $\mu$m and 70 $\mu$m catalogs we use two different approaches.
First, we compute the noise of each detection using our residual maps.
Second, we estimate a statistical flux uncertainty based on extensive Monte-Carlo simulations.
\\ \\}
\indent{
Noises estimated on residual maps correspond to the pixel dispersion, around a given source, of the residual map convolved with the appropriate PSF.
This method has the advantage of taking into account the rms of the map and the quality of our fitting procedure.
These noise estimates are given in our released catalogs as $\sigma^{map}$.
These estimates are almost equal to the rms of our maps, i.e., $\sigma^{map}\thicksim3\  \mu$Jy/beam at 24~$\mu$m in both GOODS-N and GOODS-S, $0.3$~mJy/beam at 70~$\mu$m in GOODS-N and in the deepest region of GOODS-S, and $0.45$~mJy/beam at 70~$\mu$m in the shallowest region of GOODS-S.  
\\ \\}
\indent{
In order to estimate the effect of confusion noise we performed extensive Monte-Carlo simulations.
We added artificial sources in the 24~$\mu$m and~70 $\mu$m images with a flux distribution matching approximately the measured number counts \citep[see][]{frayer_2006, papovich_2004}.
To preserve the original statistics of the image (especially the crowding properties) the numbers of artificial objects added in the image was kept small (we only added 40  sources into the 24 $\mu$m images and 4 sources into 70 $\mu$m images). 
 We then performed our source extraction method and compared the resulting photometry to the input values.
To increase the statistic, we used repeatedly the same procedure with different positions in the same field.
For each field we introduced a total of $20\,000$ artificial objects.
Results of these Monte-Carlo simulations are shown in Figure 1 of \citet{magnelli_2009} and are summarized hereafter.
\\}
\indent{
From these Monte-Carlo simulations we derive three important quantities: the photometric accuracy, the completeness and the contamination of our catalogs as function of flux density.
Completeness is define as the fraction of simulated sources extracted with a flux accuracy better than 50\%.
The contamination is defined as the fraction of simulated sources introduced with $S<2\sigma^{map}$ which are extracted with $S>3\sigma^{map}$.
\\}
\indent{
Using these Monte-Carlo simulations, we find that in both GOODS fields our 24 $\mu$m catalogs are 80$\%$ complete at $\thicksim30$ $\mu$Jy.
At this flux density, the flux accuracy is better that 20$\%$ and the contamination is $\thicksim10\%$.
The flux accuracy of our source extraction reaches 33$\%$ around 20 $\mu$Jy.
This limit could be defined as the ``real" $3\sigma^{simu}$ limit of our data because this estimate take into account confusion noise.
At 20~$\mu$Jy, the completeness of our catalog is $\thicksim40\%$ and the contamination is $\thicksim15\%$.
\\}
\indent{
For our deep 70 $\mu$m data in GOODS-N and -S, Monte-Carlo simulations show that our catalogs are 80$\%$ complete at $2.5\,$mJy.
The 33$\%$ flux accuracy is reached at $2$ mJy with a completeness of $\thicksim50\%$ and a contamination of $\thicksim15\%$.
For the shallow 70 $\mu$m data of GOODS-S, the $80\%$ completeness limit is reached at $3$ mJy, and the $33\%$ flux accuracy is reached at $2.5$~mJy.
At $2.5$ mJy, the completeness is $45\%$ and the contamination is $15\%$.
\\}
\indent{
Flux uncertainties derived using our Monte-Carlo simulations are denoted by $\sigma^{simu}$.
These flux uncertainties present the advantage of accounting for nearly all sources of noise, which explains why they are almost always larger than noise estimates based on residual maps (i.e., $\sigma^{map}$). 
However, this noise estimate is computed independently of the actual position of the individual sources, it is statistical.
In some cases, local effects can dominate the noise as it is the case when two sources are blended.
This local effect, together with the background fluctuation due to the photometric confusion noise (i.e. the noise due to sources fainter that the detection limit that were not subtracted from the image to produce the residual image), is better accounted for in the noise estimated from the residual maps, which is estimated locally.
To be conservative, users should always use the highest uncertainties between $\sigma^{map}$ and $\sigma^{simu}$, but not the quadratic combination of both since they are not independent.\\ \\}
\indent{
Tables \ref{tab:sources 1}, \ref{tab:sources 2}, \ref{tab:sources 3} and \ref{tab:sources 4} give excerpt of our complete GOODS-N/S 24 $\mu$m and 70 $\mu$m catalogs available at CDS (http://cdsweb.u-strasbg.fr/cgi-bin/qcat?J/A+A/$<$volume$>$/$<$page$>$).
For each field we decide to split our 24 $\mu$m catalogs into two (i.e., sources with $3<\sigma^{simu}<5$ and sources with $5>\sigma^{simu}$) in order to highlight that in deep and confused fields the use of sources below $5$-$\sigma$ has to be done with caution.  
Positions of the 24 $\mu$m and 70 $\mu$m sources correspond to the IRAC positions used as priors to our source extraction.
IRAC coordinates are calibrated to match the GOODS ACS version 2 coordinate system.
For 24 $\mu$m sources that are not individually detected at 70~$\mu$m, we report an upper flux limit computed from our residual maps (i.e., $5$-$\sigma^{map}$ at the position of the source).
\\}

\begin{table*}
\caption{\label{tab:fit parameter tot}Parameter values of the rest-frame 8 $\mu$m, 15 $\mu$m, 25 $\mu m$ and 35 $\mu m$ LF}
\centering
\begin{tabular}{cccccc}
\hline \hline
{Redshift} & {Wavelength} &{$\alpha_{1}\,^\mathrm{a}$} &{$\alpha_{2}\,^\mathrm{a}$} &{\rm{Log($L_{knee}$)} } & {\rm{Log($\phi_{knee}$)}} \\
& & & &  {\tiny{\rm{Log($L_{\odot}$)}}} & {\tiny{\rm{Log($Mpc^{-3}dex^{-1}$)}}} \\
\hline
$z\thicksim0$ & 8 $\mu$m & $-0.8$ & $-3.2$ &  $10.29\pm0.01 $ & $-2.92\pm0.01$\\
$1.3<z<1.8$ & 8 $\mu$m & $-0.8$ & $-3.2$ & $11.33\pm0.07$ & $-3.35\pm0.09$\\ 
$1.8<z<2.3$ & 8 $\mu$m & $-0.8$ & $-3.2$ & $11.41\pm0.06 $ & $-3.63\pm0.09$\\ \\ 
$z\thicksim0$ & 15 $\mu$m & $-0.57$ & $-2.27$ &  $9.56\pm0.04 $ & $-2.73\pm0.07$\\
$0.4<z<0.7\,^\mathrm{b}$ & 15 $\mu$m & $-0.57$ & $-2.27$ & $10.22\pm0.03$ & $-2.63\pm0.05$\\
$0.7<z<1.0\,^\mathrm{b}$ & 15 $\mu$m & $-0.57$ & $-2.27$ & $10.57\pm0.04$ & $-2.86\pm0.04$\\ 
$1.0<z<1.3\,^\mathrm{b}$ & 15 $\mu$m & $-0.57$ & $-2.27$ & $10.79\pm0.05 $ & $-2.93\pm0.06$\\
$1.3<z<1.8$ & 15 $\mu$m & $-0.57$ & $-2.27$ & $10.85\pm0.06$ & $-3.02\pm0.08$\\ 
$1.8<z<2.3$ & 15 $\mu$m & $-0.57$ & $-2.27$ & $10.99\pm0.08 $ & $-3.17\pm0.11$\\ \\ 
$z\thicksim0$ & 25 $\mu$m & $-0.6$ & $-2.2$ &  $9.91\pm0.01 $ & $-3.04\pm0.02$\\
$1.3<z<1.8$ & 25 $\mu$m & $-0.6$ & $-2.2$ & $11.12\pm0.11$ & $-3.12\pm0.13$\\ 
$1.8<z<2.3$ & 25 $\mu$m & $-0.6$ & $-2.2$ & $11.18\pm0.07 $ & $-3.14\pm0.11$\\ \\ 
$z\thicksim0$ & 35 $\mu$m & $-0.55$ & $-1.95$ & $9.85\pm0.07 $ & $-2.83\pm0.10$\\
$0.4<z<0.7\,^\mathrm{b}$ & 35 $\mu$m & $-0.55$ & $-1.95$ & $10.73\pm0.04$ & $-2.98\pm0.05$\\
$0.7<z<1.0\,^\mathrm{b}$ & 35 $\mu$m & $-0.55$ & $-1.95$ & $10.82\pm0.04$ & $-2.73\pm0.06$\\
$1.0<z<1.3\,^\mathrm{b}$ & 35 $\mu$m & $-0.55$ & $-1.95$ & $11.20\pm0.04 $ & $-2.98\pm0.05$\\
$1.3<z<1.8$ & 35 $\mu$m & $-0.55$ & $-1.95$ & $11.26\pm0.11$ & $-3.12\pm0.12$\\ 
$1.8<z<2.3$ & 35 $\mu$m & $-0.55$ & $-1.95$ & $11.34\pm0.07 $ & $-3.10\pm0.09$\\ \\ 
\hline
\end{tabular}
\begin{list}{}{}
\item[$^{\mathrm{a}}$] Fixed slopes
\item[$^{\mathrm{b}}$] These parameter values are taken from \citet{magnelli_2009}
\end{list}
\end{table*}

\begin{table*}
\caption{\label{tab:LF 8} The rest-frame 8 $\mu$m LF derived from the $1/V_{max}$ analysis}
\centering
\begin{tabular}{cccc}
\hline \hline
\multicolumn{2}{c}{\rule{0pt}{3ex}$1.3<z<1.8$} & \multicolumn{2}{c}{\rule{0pt}{3ex}$1.8<z<2.3$} \\
\hline
{${\rm log}(L_{8\,\mu\rm{m}}^{low})-{\rm log}(L_{8\,\mu\rm{m}}^{high})$} & {\rm{log($\phi$)}} & {${\rm log}(L_{8\,\mu\rm{m}}^{low})-{\rm log}(L_{8\,\mu\rm{m}}^{high})$} & {\rm{log($\phi$)}} \\
{\tiny{\rm{log($L_{\odot}$)}}} & {\tiny{\rm{log(${\rm Mpc^{-3}dex^{-1}}$)}}} & {\tiny{\rm{log($L_{\odot}$)}}} & {\tiny{\rm{log(${\rm Mpc^{-3}dex^{-1}}$)}}}\\
\hline
\rule{0pt}{3ex}10.4 - 10.6 & $-2.68^{+0.10}_{-0.10}$ & 10.6 - 10.8 & $-3.14^{+0.11}_{-0.11}$\\ 
\rule{0pt}{3ex}10.6 - 10.8 & $-2.85^{+0.10}_{-0.10}$ & 10.8 - 11.1 & $-3.16^{+0.11}_{-0.11}$\\ 
\rule{0pt}{3ex}10.8 - 11.0 & $-2.94^{+0.10}_{-0.11}$ & 11.1 - 11.4 & $-3.47^{+0.11}_{-0.12}$\\ 
\rule{0pt}{3ex}11.0 - 11.2 & $-3.12^{+0.11}_{-0.11}$ & 11.4 - 11.7 & $-3.96^{+0.14}_{-0.16}$\\ 
\rule{0pt}{3ex}11.2 - 11.6 & $-3.59^{+0.11}_{-0.12}$ & 11.7 - 12.0 & $-5.14^{+0.31}_{-5.14}$\\ 
\rule{0pt}{3ex}11.6 - 12.0 & $-4.63^{+0.20}_{-0.31}$ & \multicolumn{2}{c}{\dots} \\ 
\rule{0pt}{3ex}12.0 - 12.4 & $-5.23^{+0.31}_{-5.23}$ & \multicolumn{2}{c}{\dots} \\ \\
\hline
\end{tabular}
\end{table*}

\begin{table*}
\caption{\label{tab:LF 15} The rest-frame 15 $\mu$m LF derived from the $1/V_{max}$ analysis}
\centering
\begin{tabular}{cccc}
\hline \hline
\multicolumn{2}{c}{\rule{0pt}{3ex}$1.3<z<1.8$} & \multicolumn{2}{c}{\rule{0pt}{3ex}$1.8<z<2.3$} \\
\hline
{${\rm log}(L_{15\,\mu\rm{m}}^{low})-{\rm log}(L_{15\,\mu\rm{m}}^{high})$} & {\rm{log($\phi$)}} & {${\rm log}(L_{15\,\mu\rm{m}}^{low})-{\rm log}(L_{15\,\mu\rm{m}}^{high})$} & {\rm{log($\phi$)}} \\
{\tiny{\rm{log($L_{\odot}$)}}} & {\tiny{\rm{log(${\rm Mpc^{-3}dex^{-1}}$)}}} & {\tiny{\rm{log($L_{\odot}$)}}} & {\tiny{\rm{log(${\rm Mpc^{-3}dex^{-1}}$)}}}\\
\hline
\rule{0pt}{3ex}10.1 - 10.3 & $-2.54^{+0.10}_{-0.11}$ & 10.4 - 10.6 & $-2.77^{+0.11}_{-0.11}$\\ 
\rule{0pt}{3ex}10.3 - 10.6 & $-2.79^{+0.10}_{-0.10}$ & 10.6 - 10.9 & $-3.05^{+0.10}_{-0.11}$\\ 
\rule{0pt}{3ex}10.6 - 10.9 & $-3.01^{+0.10}_{-0.10}$ & 10.9 - 11.2 & $-3.37^{+0.11}_{-0.11}$\\ 
\rule{0pt}{3ex}10.9 - 11.2 & $-3.45^{+0.11}_{-0.12}$ & 11.2 - 11.6 & $-4.01^{+0.13}_{-0.15}$\\ 
\rule{0pt}{3ex}11.2 - 11.7 & $-4.32^{+0.15}_{-0.19}$ & 11.6 - 12.1 & $-5.06^{+0.25}_{-0.54}$\\ 
\rule{0pt}{3ex}11.7 - 12.2 & $-5.02^{+0.25}_{-0.54}$ & \multicolumn{2}{c}{\dots} \\ \\
\hline
\end{tabular}
\end{table*}
\begin{table*}
\caption{\label{tab:LF 25} The rest-frame 25 $\mu$m LF derived from the $1/V_{max}$ analysis}
\centering
\begin{tabular}{cccc}
\hline \hline
\multicolumn{2}{c}{\rule{0pt}{3ex}$1.3<z<1.8$} & \multicolumn{2}{c}{\rule{0pt}{3ex}$1.8<z<2.3$} \\
\hline
{${\rm log}(L_{25\,\mu\rm{m}}^{low})-{\rm log}(L_{25\,\mu\rm{m}}^{high})$} & {\rm{log($\phi$)}} & {${\rm log}(L_{25\,\mu\rm{m}}^{low})-{\rm log}(L_{25\,\mu\rm{m}}^{high})$} & {\rm{log($\phi$)}} \\
{\tiny{\rm{log($L_{\odot}$)}}} & {\tiny{\rm{log(${\rm Mpc^{-3}dex^{-1}}$)}}} & {\tiny{\rm{log($L_{\odot}$)}}} & {\tiny{\rm{log(${\rm Mpc^{-3}dex^{-1}}$)}}}\\
\hline
\rule{0pt}{3ex}10.3 - 10.6 & $-2.65^{+0.10}_{-0.10}$ & 10.6 - 11.0 & $-2.90^{+0.10}_{-0.10}$\\ 
\rule{0pt}{3ex}10.6 - 10.9 & $-2.82^{+0.10}_{-0.10}$ & 11.0 - 11.4 & $-3.17^{+0.10}_{-0.10}$\\ 
\rule{0pt}{3ex}10.9 - 11.2 & $-3.21^{+0.11}_{-0.11}$ & 11.4 - 11.8 & $-4.06^{+0.13}_{-0.16}$\\ 
\rule{0pt}{3ex}11.2 - 11.7 & $-3.74^{+0.11}_{-0.12}$ & 11.8 - 12.2 & $-4.97^{+0.25}_{-0.54}$\\ 
\rule{0pt}{3ex}11.7 - 12.2 & $-4.85^{+0.22}_{-0.38}$ & \multicolumn{2}{c}{\dots} \\ \\
\hline
\end{tabular}
\end{table*}
\begin{table*}
\caption{\label{tab:LF 35} The rest-frame 35 $\mu$m LF derived from the $1/V_{max}$ analysis}
\centering
\begin{tabular}{cccc}
\hline \hline
\multicolumn{2}{c}{\rule{0pt}{3ex}$1.3<z<1.8$} & \multicolumn{2}{c}{\rule{0pt}{3ex}$1.8<z<2.3$} \\
\hline
{${\rm log}(L_{35\,\mu\rm{m}}^{low})-{\rm log}(L_{35\,\mu\rm{m}}^{high})$} & {\rm{log($\phi$)}} & {${\rm log}(L_{35\,\mu\rm{m}}^{low})-{\rm log}(L_{35\,\mu\rm{m}}^{high})$} & {\rm{log($\phi$)}} \\
{\tiny{\rm{log($L_{\odot}$)}}} & {\tiny{\rm{log(${\rm Mpc^{-3}dex^{-1}}$)}}} & {\tiny{\rm{log($L_{\odot}$)}}} & {\tiny{\rm{log(${\rm Mpc^{-3}dex^{-1}}$)}}}\\
\hline
\rule{0pt}{3ex}10.5 - 10.8 & $-2.78^{+0.10}_{-0.10}$ & 10.7 - 11.0 & $-2.80^{+0.11}_{-0.11}$\\ 
\rule{0pt}{3ex}10.8 - 11.1 & $-2.83^{+0.10}_{-0.10}$ & 11.0 - 11.2 & $-2.95^{+0.11}_{-0.11}$\\ 
\rule{0pt}{3ex}11.1 - 11.4 & $-3.22^{+0.11}_{-0.11}$ & 11.2 - 11.6 & $-3.16^{+0.10}_{-0.10}$\\ 
\rule{0pt}{3ex}11.4 - 11.9 & $-3.76^{+0.12}_{-0.12}$ & 11.6 - 12.1 & $-3.96^{+0.12}_{-0.13}$\\ 
\rule{0pt}{3ex}11.9 - 12.4 & $-4.85^{+0.22}_{-0.38}$ & 12.1 - 12.6 & $-5.36^{+0.31}_{-5.36}$\\  \\ 
\hline
\end{tabular}
\end{table*}
\begin{table*}
\caption{\label{tab:LF IR} The infrared LF derived from the $1/V_{max}$ analysis using $L_{IR}^{fit}$}
\centering
\begin{tabular}{cccccccc}
\hline \hline
\multicolumn{2}{c}{\rule{0pt}{3ex}$1.3<z<1.8$} & \multicolumn{2}{c}{\rule{0pt}{3ex}$1.8<z<2.3$} \\
\hline
{${\rm log}(L_{IR}^{low})-{\rm log}(L_{IR}^{high})$} & {\rm{log($\phi$)}} & {${\rm log}(L_{IR}^{low})-{\rm log}(L_{IR}^{high})$} & {\rm{log($\phi$)}} \\
{\tiny{\rm{log($L_{\odot}$)}}} & {\tiny{\rm{log(${\rm Mpc^{-3}dex^{-1}}$)}}} & {\tiny{\rm{log($L_{\odot}$)}}} & {\tiny{\rm{log(${\rm Mpc^{-3}dex^{-1}}$)}}}\\
\hline
\rule{0pt}{3ex}10.8 - 11.2 & $-2.38^{+0.25}_{-0.25}$ & 11.25 - 11.55 & $-2.72^{+0.25}_{-0.25}$\\ 
\rule{0pt}{3ex}11.2 - 11.6 & $-2.78^{+0.25}_{-0.25}$ & 11.55 - 11.85 & $-3.05^{+0.25}_{-0.25}$\\ 
\rule{0pt}{3ex}11.6 - 12.0 & $-3.15^{+0.25}_{-0.25}$ & 11.85 - 12.15 & $-3.29^{+0.25}_{-0.25}$\\ 
\rule{0pt}{3ex}12.0 - 12.4 & $-3.69^{+0.26}_{-0.26}$ & 12.15 - 12.45 & $-3.88^{+0.26}_{-0.27}$\\ 
\rule{0pt}{3ex}12.4 - 12.8 & $-4.75^{+0.31}_{-0.45}$ & 12.45- 12.75 & $-4.84^{+0.34}_{-0.58}$\\
 \multicolumn{2}{c}{\dots} &  \rule{0pt}{3ex} 12.75 - 13.05 & $-5.14^{+0.39}_{-5.14}$\\ \\
\hline
\end{tabular}
\end{table*}
\begin{table*}
\caption{\label{tab:LF IR bis} The infrared LF derived from the $1/V_{max}$ analysis using $L_{IR}^{70}$}
\centering
\begin{tabular}{cccccccc}
\hline \hline
\multicolumn{2}{c}{\rule{0pt}{3ex}$1.3<z<1.8$} & \multicolumn{2}{c}{\rule{0pt}{3ex}$1.8<z<2.3$} \\
\hline
{${\rm log}(L_{IR}^{low})-{\rm log}(L_{IR}^{high})$} & {\rm{log($\phi$)}} & {${\rm log}(L_{IR}^{low})-{\rm log}(L_{IR}^{high})$} & {\rm{log($\phi$)}} \\
{\tiny{\rm{log($L_{\odot}$)}}} & {\tiny{\rm{log(${\rm Mpc^{-3}dex^{-1}}$)}}} & {\tiny{\rm{log($L_{\odot}$)}}} & {\tiny{\rm{log(${\rm Mpc^{-3}dex^{-1}}$)}}}\\
\hline
\rule{0pt}{3ex}10.8 - 11.4 & $-2.58^{+0.25}_{-0.25}$ & 11.15 - 11.45 & $-2.74^{+0.25}_{-0.25}$\\ 
\rule{0pt}{3ex}11.4 - 12.0 & $-3.11^{+0.25}_{-0.25}$ & 11.45 - 11.75 & $-3.19^{+0.25}_{-0.25}$\\ 
\rule{0pt}{3ex}12.0 - 12.6 & $-3.91^{+0.26}_{-0.26}$ & 11.75 - 12.05 & $-3.11^{+0.25}_{-0.25}$\\ 
\rule{0pt}{3ex}12.6 - 13.2 & $-5.40^{+0.39}_{-5.40}$ & 12.05 - 12.35 & $-3.76^{+0.26}_{-0.26}$\\ 
 \multicolumn{2}{c}{\dots}                       & \rule{0pt}{3ex}12.35 - 12.65 & $-4.84^{+0.34}_{-0.58}$\\ \\
\hline
\end{tabular}
\end{table*}

\begin{table*}
\caption{\label{tab:sources 1} MIPS sources in GOODS-N with $S_{24\,\mu{\rm m}}/\sigma^{simu}_{24 \mu{\rm m}}>5$.
The second and third columns list the prior position from IRAC.
Fourth column lists the 24 $\mu$m flux density.
Fifth and sixth columns give the 24 $\mu$m flux uncertainty derived from the residual map and from our Monte-Carlo simulations respectively (see text for detail).
Seventh column indicates the integration time on the source.
Columns 8 to 11 repeat columns 4-7 for 70 $\mu$m.
}
\begin{tabular}{cccccccccc}
\hline\hline
Name & IRAC position & $F_{24 \mu{\rm m}}$ & $\sigma_{24 \mu{\rm m}}^{Map}$ & $\sigma_{24 \mu{\rm m}}^{Simu}$ & $Cov_{24 \mu{\rm m}}$ & $F_{70 \mu{\rm m}}$ & $\sigma_{70 \mu{\rm m}}^{Map}$ & $\sigma_{70 \mu{\rm m}}^{Simu}$ & $Cov_{70 \mu{\rm m}}$ \\
&{\tiny$\alpha_{{\rm J2000}}$\hspace{1cm}$\delta_{{\rm J2000}}$} & {\tiny $\mu$Jy} & {\tiny $\mu$Jy}& {\tiny $\mu$Jy} & {\tiny s} & {\tiny mJy} & {\tiny mJy}& {\tiny mJy}  & {\tiny s}\\
\hline
   MIPSJ123539.5+621129.0  &  12:35:39.48  +62:11:29.02  &  $      42.6  $  &  $       5.6  $  &  $       6.7  $  &  $ 10873  $  &  $< 2.9$  &  \dots  &  \dots$  $  &  $  5601 $ \\
   MIPSJ123539.5+621243.8  &  12:35:39.54  +62:12:43.78  &  $      54.0  $  &  $       4.6  $  &  $       6.7  $  &  $ 10239  $  &  $< 2.2$  &  \dots  &  \dots$  $  &  $  4747 $ \\
   MIPSJ123539.9+621324.8  &  12:35:39.94  +62:13:24.79  &  $      52.2  $  &  $       4.2  $  &  $       6.7  $  &  $  9512  $  &  $< 2.4$  &  \dots  &  \dots$  $  &  $  3728 $ \\
   MIPSJ123540.2+621224.2  &  12:35:40.17  +62:12:24.16  &  $     107.1  $  &  $       7.2  $  &  $       8.2  $  &  $ 11518  $  &  $< 3.5$  &  \dots  &  \dots$  $  &  $  5453 $ \\
   MIPSJ123540.7+621218.9  &  12:35:40.70  +62:12:18.93  &  $     107.1  $  &  $       5.8  $  &  $       8.2  $  &  $ 11934  $  &  $< 3.5$  &  \dots  &  \dots$  $  &  $  5711 $ \\
   MIPSJ123541.0+621136.1  &  12:35:41.00  +62:11:36.13  &  $     139.8  $  &  $       4.1  $  &  $       8.3  $  &  $ 12437  $  &  $< 4.6$  &  \dots  &  \dots$  $  &  $  6409 $ \\
   MIPSJ123541.4+621217.4  &  12:35:41.39  +62:12:17.37  &  $     419.5  $  &  $       5.2  $  &  $      11.7  $  &  $ 12395  $  &  $       2.6  $  &  $       0.3  $  &  $       0.6  $  &  $  6102 $ \\
   MIPSJ123541.4+621316.5  &  12:35:41.41  +62:13:16.51  &  $      41.3  $  &  $       4.1  $  &  $       6.6  $  &  $ 12733  $  &  $< 1.6$  &  \dots  &  \dots$  $  &  $  4483 $ \\
   MIPSJ123541.6+621151.0  &  12:35:41.60  +62:11:51.01  &  $      44.7  $  &  $       3.5  $  &  $       6.7  $  &  $ 12662  $  &  $< 4.2$  &  \dots  &  \dots$  $  &  $  6771 $ \\
   MIPSJ123541.7+621223.5  &  12:35:41.73  +62:12:23.54  &  $      64.0  $  &  $       5.2  $  &  $       7.0  $  &  $ 12853  $  &  $< 2.4$  &  \dots  &  \dots$  $  &  $  6014 $ \\
     \hline
\end{tabular}
\end{table*}

\begin{table*}
\caption{\label{tab:sources 2} MIPS sources in GOODS-N with $3<S_{24\,\mu{\rm m}}/\sigma^{simu}_{24 \mu{\rm m}}<5$.
Columns are the same as in Table \ref{tab:sources 1}.
}
\begin{tabular}{cccccccccc}
\hline\hline
Name & IRAC position & $F_{24 \mu{\rm m}}$ & $\sigma_{24 \mu{\rm m}}^{Map}$ & $\sigma_{24 \mu{\rm m}}^{Simu}$ & $Cov_{24 \mu{\rm m}}$ & $F_{70 \mu{\rm m}}$ & $\sigma_{70 \mu{\rm m}}^{Map}$ & $\sigma_{70 \mu{\rm m}}^{Simu}$ & $Cov_{70 \mu{\rm m}}$ \\
&{\tiny$\alpha_{{\rm J2000}}$\hspace{1cm}$\delta_{{\rm J2000}}$} & {\tiny $\mu$Jy} & {\tiny $\mu$Jy}& {\tiny $\mu$Jy} & {\tiny s} & {\tiny mJy} & {\tiny mJy}& {\tiny mJy}  & {\tiny s}\\
\hline
   MIPSJ123540.2+621108.2  &  12:35:40.23  +62:11:08.16  &  $      23.8  $  &  $       4.2  $  &  $       6.8  $  &  $ 11885  $  &  $< 2.0$  &  \dots  &  \dots$  $  &  $  5370 $ \\
   MIPSJ123541.3+621047.2  &  12:35:41.29  +62:10:47.22  &  $      23.5  $  &  $       3.7  $  &  $       6.8  $  &  $ 11294  $  &  $< 3.4$  &  \dots  &  \dots$  $  &  $  5363 $ \\
   MIPSJ123543.8+621218.8  &  12:35:43.77  +62:12:18.82  &  $      23.5  $  &  $       3.5  $  &  $       6.8  $  &  $ 15118  $  &  $< 2.4$  &  \dots  &  \dots$  $  &  $  7117 $ \\
   MIPSJ123544.7+621246.8  &  12:35:44.69  +62:12:46.78  &  $      20.5  $  &  $       5.7  $  &  $       6.7  $  &  $ 19859  $  &  $< 1.3$  &  \dots  &  \dots$  $  &  $  6439 $ \\
   MIPSJ123545.2+621151.9  &  12:35:45.18  +62:11:51.91  &  $      23.3  $  &  $       5.4  $  &  $       6.8  $  &  $ 15912  $  &  $< 3.0$  &  \dots  &  \dots$  $  &  $  7764 $ \\
   MIPSJ123545.3+621134.0  &  12:35:45.30  +62:11:33.97  &  $      20.5  $  &  $       4.5  $  &  $       6.7  $  &  $ 16640  $  &  $< 4.6$  &  \dots  &  \dots$  $  &  $  7482 $ \\
   MIPSJ123545.4+621306.4  &  12:35:45.44  +62:13:06.39  &  $      31.6  $  &  $       3.7  $  &  $       6.6  $  &  $ 18447  $  &  $< 1.5$  &  \dots  &  \dots$  $  &  $  6228 $ \\
   MIPSJ123545.6+621034.7  &  12:35:45.63  +62:10:34.68  &  $      29.3  $  &  $       7.1  $  &  $       6.6  $  &  $ 13190  $  &  $< 3.1$  &  \dots  &  \dots$  $  &  $  6388 $ \\
   MIPSJ123547.6+621147.2  &  12:35:47.59  +62:11:47.18  &  $      31.2  $  &  $       7.2  $  &  $       6.7  $  &  $ 18975  $  &  $< 3.5$  &  \dots  &  \dots$  $  &  $  8395 $ \\
   MIPSJ123550.3+621423.6  &  12:35:50.33  +62:14:23.64  &  $      21.7  $  &  $       4.2  $  &  $       6.8  $  &  $ 13135  $  &  $< 3.8$  &  \dots  &  \dots$  $  &  $  4727 $ \\
     \hline
\end{tabular}
\end{table*}

\begin{table*}
\caption{\label{tab:sources 3} MIPS sources in GOODS-S with $S_{24\,\mu{\rm m}}/\sigma^{simu}_{24 \mu{\rm m}}>5$.
Columns are the same as in Table \ref{tab:sources 1}.
}
\begin{tabular}{cccccccccc}
\hline\hline
Name & IRAC position & $F_{24 \mu{\rm m}}$ & $\sigma_{24 \mu{\rm m}}^{Map}$ & $\sigma_{24 \mu{\rm m}}^{Simu}$ & $Cov_{24 \mu{\rm m}}$ & $F_{70 \mu{\rm m}}$ & $\sigma_{70 \mu{\rm m}}^{Map}$ & $\sigma_{70 \mu{\rm m}}^{Simu}$ & $Cov_{70 \mu{\rm m}}$ \\
&{\tiny$\alpha_{{\rm J2000}}$\hspace{1cm}$\delta_{{\rm J2000}}$} & {\tiny $\mu$Jy} & {\tiny $\mu$Jy}& {\tiny $\mu$Jy} & {\tiny s} & {\tiny mJy} & {\tiny mJy}& {\tiny mJy}  & {\tiny s}\\
\hline
   MIPSJ033201.1-274331.1  &  03:32:01.11  -27:43:31.07  &  $     232.2  $  &  $       3.7  $  &  $       8.7  $  &  $ 17558  $  &  $< 2.2$  &  \dots  &  \dots$  $  &  $  7946 $ \\
   MIPSJ033201.2-274636.0  &  03:32:01.15  -27:46:35.98  &  $     200.7  $  &  $       6.2  $  &  $       8.1  $  &  $ 10241  $  &  $< 1.7$  &  \dots  &  \dots$  $  &  $  8105 $ \\
   MIPSJ033201.2-274134.6  &  03:32:01.24  -27:41:34.57  &  $     186.0  $  &  $       5.8  $  &  $       7.7  $  &  $ 12533  $  &  $< 2.3$  &  \dots  &  \dots$  $  &  $  6676 $ \\
   MIPSJ033201.3-274553.8  &  03:32:01.26  -27:45:53.75  &  $      72.3  $  &  $       4.4  $  &  $       7.4  $  &  $ 12171  $  &  $< 2.2$  &  \dots  &  \dots$  $  &  $  8099 $ \\
   MIPSJ033201.4-274646.5  &  03:32:01.40  -27:46:46.54  &  $     167.3  $  &  $       8.0  $  &  $       7.9  $  &  $ 10220  $  &  $< 1.4$  &  \dots  &  \dots$  $  &  $  8200 $ \\
   MIPSJ033201.5-274138.7  &  03:32:01.46  -27:41:38.71  &  $     208.1  $  &  $       6.0  $  &  $       8.2  $  &  $ 12573  $  &  $       2.8  $  &  $       0.2  $  &  $       0.8  $  &  $  6689 $ \\
   MIPSJ033201.5-274229.8  &  03:32:01.48  -27:42:29.80  &  $      95.7  $  &  $       4.5  $  &  $       8.1  $  &  $ 20170  $  &  $< 5.5$  &  \dots  &  \dots$  $  &  $  7206 $ \\
   MIPSJ033201.5-274402.5  &  03:32:01.51  -27:44:02.45  &  $      57.1  $  &  $       4.3  $  &  $       6.7  $  &  $ 17091  $  &  $< 2.3$  &  \dots  &  \dots$  $  &  $  8299 $ \\
   MIPSJ033201.6-274326.9  &  03:32:01.61  -27:43:26.93  &  $      55.4  $  &  $       2.8  $  &  $       6.8  $  &  $ 19023  $  &  $< 1.8$  &  \dots  &  \dots$  $  &  $  7935 $ \\
   MIPSJ033201.7-274349.4  &  03:32:01.66  -27:43:49.37  &  $      51.0  $  &  $       4.1  $  &  $       6.7  $  &  $ 18518  $  &  $< 1.8$  &  \dots  &  \dots$  $  &  $  8249 $ \\
 \hline
\end{tabular}
\end{table*}

\begin{table*}
\caption{\label{tab:sources 4} MIPS sources in GOODS-S with $3<S_{24\,\mu{\rm m}}/\sigma^{simu}_{24 \mu{\rm m}}<5$.
Columns are the same as in Table \ref{tab:sources 1}.
}
\begin{tabular}{cccccccccc}
\hline\hline
Name & IRAC position & $F_{24 \mu{\rm m}}$ & $\sigma_{24 \mu{\rm m}}^{Map}$ & $\sigma_{24 \mu{\rm m}}^{Simu}$ & $Cov_{24 \mu{\rm m}}$ & $F_{70 \mu{\rm m}}$ & $\sigma_{70 \mu{\rm m}}^{Map}$ & $\sigma_{70 \mu{\rm m}}^{Simu}$ & $Cov_{70 \mu{\rm m}}$ \\
&{\tiny$\alpha_{{\rm J2000}}$\hspace{1cm}$\delta_{{\rm J2000}}$} & {\tiny $\mu$Jy} & {\tiny $\mu$Jy}& {\tiny $\mu$Jy} & {\tiny s} & {\tiny mJy} & {\tiny mJy}& {\tiny mJy}  & {\tiny s}\\
\hline
   MIPSJ033158.1-274207.9  &  03:31:58.08  -27:42:07.92  &  $      23.5  $  &  $       5.0  $  &  $       6.8  $  &  $ 10614  $  &  $< 2.3$  &  \dots  &  \dots$  $  &  $  6845 $ \\
   MIPSJ033158.9-274359.1  &  03:31:58.93  -27:43:59.09  &  $      27.2  $  &  $       5.3  $  &  $       6.8  $  &  $ 10417  $  &  $< 1.5$  &  \dots  &  \dots$  $  &  $  7976 $ \\
   MIPSJ033159.1-274421.1  &  03:31:59.12  -27:44:21.14  &  $      29.8  $  &  $       4.6  $  &  $       6.7  $  &  $  9963  $  &  $< 2.1$  &  \dots  &  \dots$  $  &  $  8036 $ \\
   MIPSJ033159.2-274145.8  &  03:31:59.18  -27:41:45.82  &  $      22.5  $  &  $       6.9  $  &  $       6.9  $  &  $ 10825  $  &  $< 1.5$  &  \dots  &  \dots$  $  &  $  6791 $ \\
   MIPSJ033159.7-274459.8  &  03:31:59.71  -27:44:59.82  &  $      21.7  $  &  $       6.8  $  &  $       6.8  $  &  $  9978  $  &  $< 2.9$  &  \dots  &  \dots$  $  &  $  8135 $ \\
   MIPSJ033200.2-274526.6  &  03:32:00.22  -27:45:26.57  &  $      20.9  $  &  $       6.3  $  &  $       6.8  $  &  $ 10371  $  &  $< 3.9$  &  \dots  &  \dots$  $  &  $  8168 $ \\
   MIPSJ033200.3-274542.4  &  03:32:00.31  -27:45:42.43  &  $      28.8  $  &  $       5.3  $  &  $       6.6  $  &  $  9808  $  &  $< 1.9$  &  \dots  &  \dots$  $  &  $  7943 $ \\
   MIPSJ033200.6-274521.6  &  03:32:00.60  -27:45:21.61  &  $      26.7  $  &  $       5.6  $  &  $       6.6  $  &  $ 11729  $  &  $< 4.5$  &  \dots  &  \dots$  $  &  $  8275 $ \\
   MIPSJ033200.8-274514.6  &  03:32:00.75  -27:45:14.60  &  $      27.0  $  &  $       4.8  $  &  $       6.7  $  &  $ 12366  $  &  $< 5.3$  &  \dots  &  \dots$  $  &  $  8385 $ \\
   MIPSJ033200.9-274408.7  &  03:32:00.90  -27:44:08.71  &  $      22.9  $  &  $       4.6  $  &  $       6.8  $  &  $ 15533  $  &  $< 2.5$  &  \dots  &  \dots$  $  &  $  8134 $ \\
\hline
\end{tabular}
\end{table*}

\end{appendix}
\end{document}